\documentclass[aps,prev,twocolumn,preprintnumbers,floatfix,nofootinbib]{revtex4-1}

\usepackage{graphicx}
\usepackage[english]{babel}
\usepackage{bm}
\usepackage{times}
\usepackage{slashed}
\usepackage{mathtools}
\usepackage{color}
\usepackage{epsfig}
\usepackage{dcolumn}
\usepackage{textcomp}
\usepackage{color}

\def\hmath$#1${\texorpdfstring{{\rmfamily\textit{#1}}}{#1}}
\usepackage{hyperref}
\usepackage{amsfonts}
\usepackage{threeparttable}

\begin{document} 

\preprint{IIPDM-2019}

\title{Doubly Charged Scalar at the High-Luminosity and High-Energy LHC}

\author{Tessio B. de Melo$^{1,2}$}
\email{tessiomelo@gmail.com}
\author{Farinaldo S. Queiroz$^1$}
\email{farinaldo.queiroz@iip.ufrn.br}
\author{Yoxara Villamizar$^1$}

\affiliation{$^1$International Institute of Physics, Universidade Federal do Rio Grande do Norte, Campus Universit\'ario, Lagoa Nova, Natal-RN 59078-970, Brazil}
\affiliation{$^2$Departamento de F\'isica, Universidade Federal da Para\'iba, Caixa Postal 5008, 58051-970, Jo\~ao Pessoa, PB, Brazil}

\begin{abstract}
Doubly charged scalars are common figures in several beyond the Standard Model studies, especially those related to neutrino masses. In this work, we estimate the High-Luminosity (HL-LHC) and High-Energy LHC (HE-LHC) sensitivity to doubly charged scalars assuming they decay promptly and exclusively into charged leptons. Our study focuses on the fit to the same-sign dilepton mass spectra and it is based on proton-proton collisions at $13$~TeV, $14$~TeV and $27$~TeV with integrated luminosity of $\mathcal{L}=139 fb^{-1}, 3 $ab$^{-1}$ and $15$ab$^{-1}$. We find that HL-LHC may probe doubly charged scalars masses up to $2.3$~TeV, whereas HE-LHC can impressively probe masses up to $3$~TeV, conclusively constituting a complementary and important probe to signs of doubly charged scalars in lepton flavor violation decays and lepton-lepton colliders.
\end{abstract}

\maketitle
\flushbottom

\section{Introduction}
\label{introduction}

Doubly charged scalars are quite popular because they are present in many beyond the Standard Model studies such as the type II seesaw mechanism, which adds a $SU(2)$ scalar triplet to the SM spectrum and represents one of the most common ways to explain neutrino masses. Such scalar triplet features a doubly charged scalar\cite{Muhlleitner:2003me,Akeroyd:2005gt,Hektor:2007uu,Perez:2008zc,Chaudhuri:2013xoa,Lindner:2016bgg,Primulando:2019evb} which is entitled to interesting phenomenological signatures in the context of lepton flavor violation. Moreover, many models in the literature based on the Left-Right \cite{Pati:1974yy,Mohapatra:1974hk,Senjanovic:1975rk,Dutta:2014dba,Dev:2016vle,Borah:2016hqn,Dev:2018kpa} and 3-3-1 symmetries \cite{CiezaMontalvo:2006zt,Alves:2011kc,Alves:2012yp,Machado:2018sfh,CarcamoHernandez:2019vih,Ferreira:2019qpf} also have this doubly charged scalar in their spectra. Other extended sectors such as Higgs triplets \cite{Gunion:1989ci,Akeroyd:2012ms,deMedeirosVarzielas:2017glw,Ghosh:2018jpa,Chala:2018opy,Chabab:2018ert,Chakraborty:2019uxk}, little higgs \cite{ArkaniHamed:2002qx,Hektor:2007uu}, Georgi-Machacek model \cite{Georgi:1985nv,Sun:2017mue,Chiang:2018cgb}, and other alternative attempts to explain neutrino masses commonly advocate for the existence of doubly charged scalars \cite{Zee:1985id,Nebot:2007bc,Camargo:2018uzw}. In summary, doubly charged scalars are indeed present in a wealth of beyond the Standard Model studies. 

For this reason, several collider searches have been conducted for such particles \cite{Mitra:2016wpr,Du:2018eaw}, specially at the LHC \cite{Huitu:1996su,Han:2007bk,Chao:2008mq,Akeroyd:2010ip,Chiang:2012dk,delAguila:2013mia,Kanemura:2013vxa,King:2014uha,kang:2014jia,Kanemura:2014ipa,Bambhaniya:2015wna}. The latter relies on the search for two prompt, isolated, and highly energetic same sign leptons. Such signal is rare in the Standard Model but it may happen with a higher rate in the aforementioned models due to the presence of the doubly charged scalar particle. The invariant mass of the same sign leptons is a good handle to discriminate potential SM background events from those stemming from the doubly charged scalar prompt decay.

\begin{figure}
    \centering
    \includegraphics[scale=0.4]{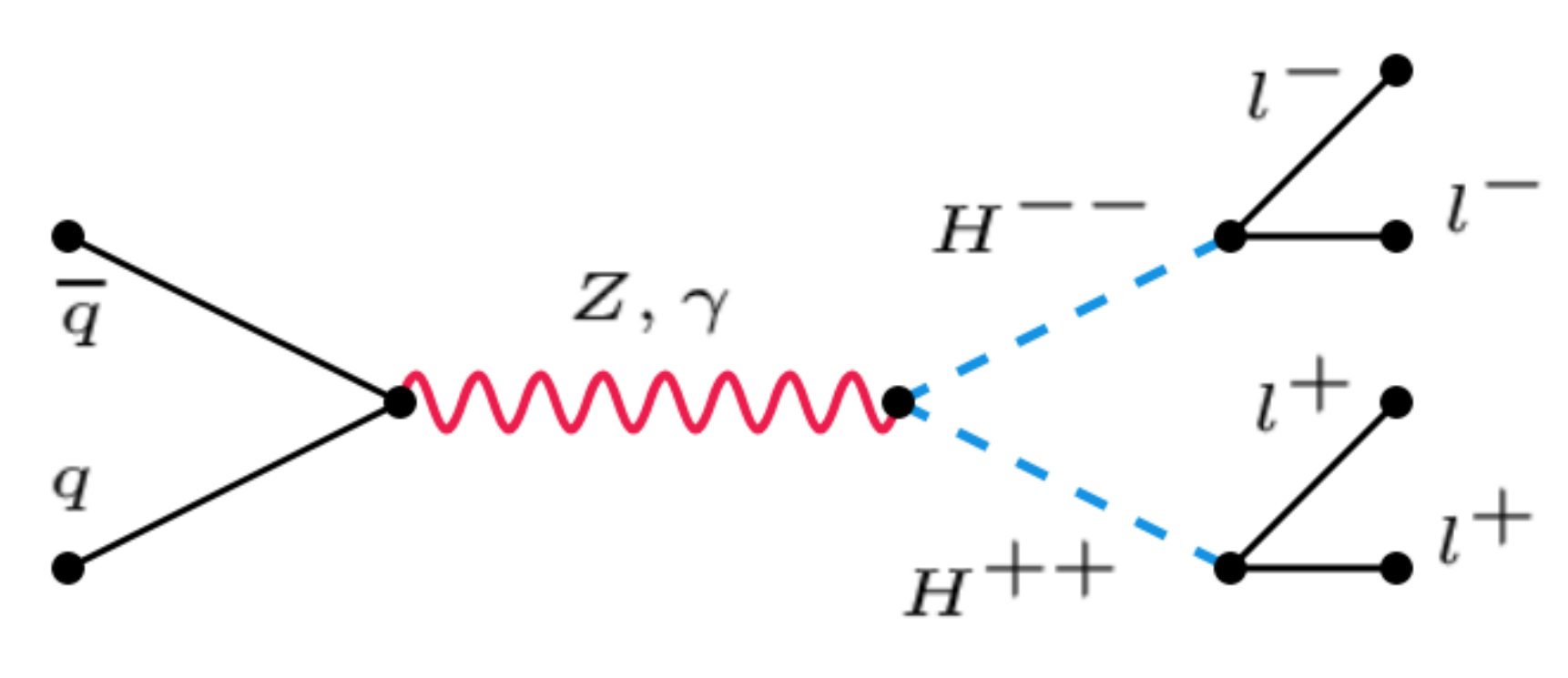}
    \caption{Feynman diagram of the resonant pair production process of the doubly charged scalar decaying into charged lepton pairs. There are other processes that contribute to the doubly charged scalar production not shown in the figure.}
    \label{diagram1}
\end{figure}

No positive signal has been observed, which led to the derivation of stringent limits with $95\%$ confidence level on the mass of the doubly charged scalar. These bounds depend on the nature of the doubly charged interactions and decays. These scalars can couple to left-handed and right-handed fermions, as occurs in left-right models. The production cross-section when couplings to left-handed leptons are involved is roughly two times larger because of the interactions involving the Z boson. We highlight that even in such models the doubly charged scalar decays mostly into charged leptons as long as the vacuum expectation value of the scalar triplet is sufficiently small. A similar conclusion applies to the canonical type II seesaw model \cite{Perez:2008ha}. In 3-3-1 models, for instance, such doubly charged scalars decay nearly exclusively into charged leptons as well. Therefore, taking the branching ratio into charged lepton to be nearly equal to unity seems a good assumption. That said, both CMS and ATLAS collaborations have performed their search for doubly charged scalars based on the pair production of the doubly charged scalar via the resonant production (see Fig.\ref{diagram1}) \cite{CMS:2012kua,ATLAS:2012hi,Chatrchyan:2012ya,ATLAS:2014kca,ATLAS:2016pbt,CMS:2016cpz,CMS:2017pet,Aaboud:2017qph} assuming the scalar decays exclusively into charged leptons. Different setups where the doubly charged scalar decays into additional particles can be easily accounted for by rescaling the bounds. An additional assumption made is that the decay width is small compared to the detector resolution, in other words, the narrow width approximation is applied. It is important to note that this is not always valid because, in models where the doubly charged scalar appears as a singlet field under $SU(2)_L$, the width can be large. Moreover, in photon initiated processes which lead to the pair production of doubly charged scalars mediated by a doubly charged scalar in the t-channel can be also important especially in the large width case \cite{Babu:2016rcr,Crivellin:2018ahj}. The inclusion of such effect might improve the LHC sensitivity to doubly charged scalars, but on the other side includes model dependent parameters. In this work, we will adopt the approach done by ATLAS and CMS collaboration which will render our study conservative.

Motivated by the LHC upgrade to High-Luminosity and High-Energy modes we obtain the sensitivity to doubly charged scalars over a wide range of decay modes.  Assuming the branching ratio into charged leptons to be 100\%, $BR(H^{\pm\pm}\rightarrow l^{\pm}l^{\pm})=100\%$, we vary the branching ratio into individual charged lepton flavors from 0\% to 100\% to assess the impact on the lower mass bounds. In agreement with \cite{Li:2018jns} we notice that as we lower the branching ratio into $ee$ pairs the lower mass bound improves because of the misidentification effect which is quite relevant for electrons with high transverse momentum \cite{Mondal:2016czu,Queiroz:2016qmc}. In particular, we consider the scenarios with $13$~TeV, $14$~TeV and $27$~TeV center of mass energy and several integrated luminosity setups as follows:
\begin{itemize}
    \item $E_{cm} =13$~TeV, $\mathcal{L}=139$~fb$^{-1}$ 
    \item $E_{cm} =14$~TeV, $\mathcal{L}=3$~ab$^{-1}$
    \item $E_{cm} =14$~TeV,$\mathcal{L}=15$~ab$^{-1}$
    \item $E_{cm} =27$~TeV, $\mathcal{L}=3$~ab$^{-1}$
    \item $E_{cm} =27$~TeV, $\mathcal{L}=15$~ab$^{-1}$
\end{itemize}

We obtain lower mass bounds for fifty different decay configurations for each collider setup outlined above. In summary, we aim with this work to provide HL-LHC and HE-LHC reach to the doubly charged scalars covering several different decays modes. We start in section II discussing the data set and our model assumptions. In section III we present our bounds and later draw our conclusions. 

\section{Data set}

Before forecasting the HL-LHC and HE-LHC sensitivity we attempted to reproduce the CMS and ATLAS results for a center of energy of $13$~TeV with an integrated luminosity of $\mathcal{L}=36.1$~fb$^{-1}$ \cite{Aaboud:2017qph}. We followed the recipe described in \cite{Aaboud:2017qph} and required each electron in the dilepton events to have at least $30$~GeV of transfer momentum ($p_T$). We applied a similar cut for dimuon events. The same logic applies to events having both flavors.  Concerning the angular distribution, only electron candidates with a pseudo-rapidity $|\eta| < 2.47$ are selected because outside of this region the calorimeter features poor reconstruction. Muon candidates identified in the muon spectrometer are required to have  $|\eta| < 2.5$ \cite{Aad:2016jkr}.

\begin{figure}[!t]
    \includegraphics[scale=0.5]{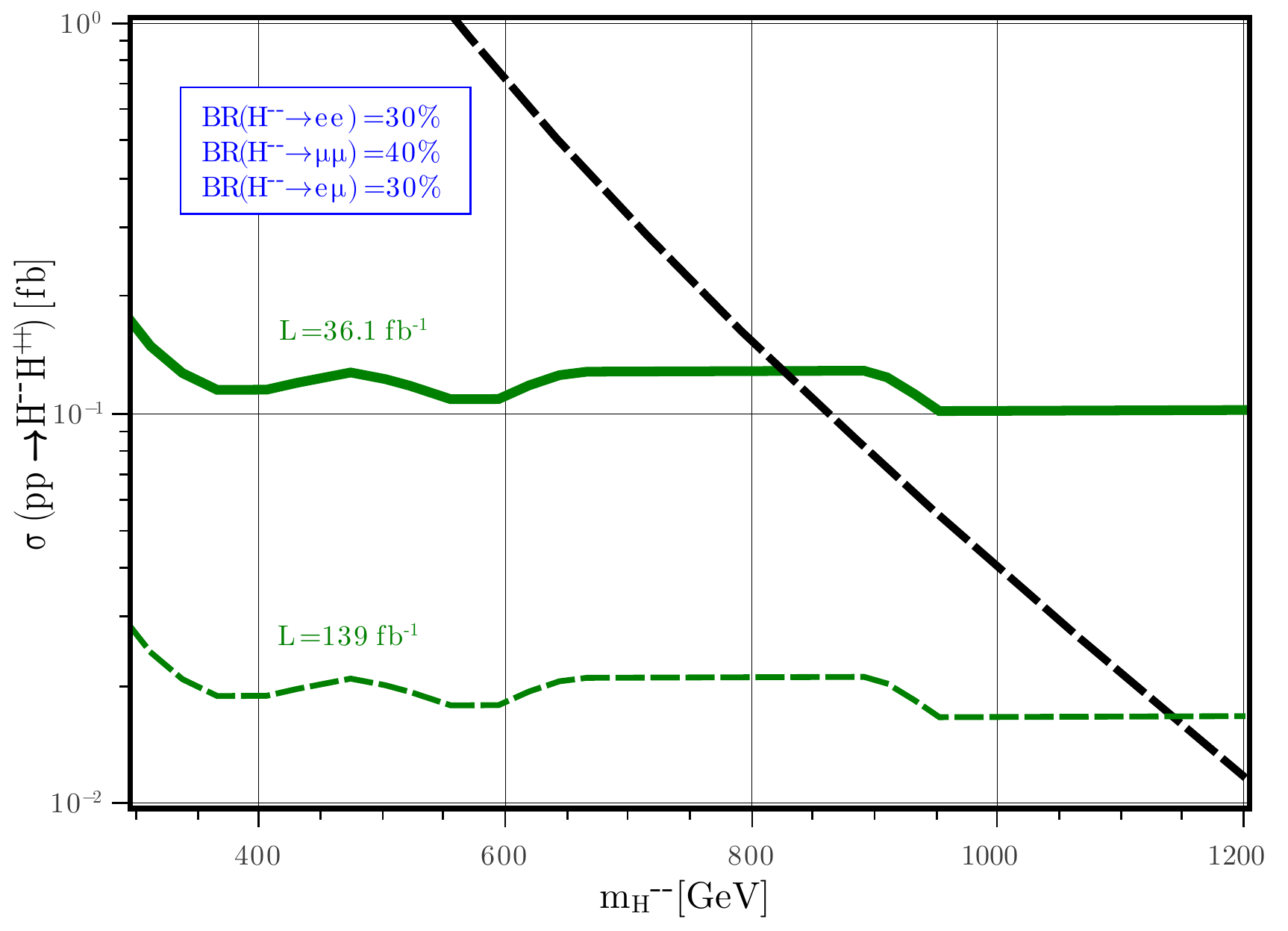}
    \caption{Production cross section as a function of the doubly charged scalar mass assuming $BR (H^{--}\rightarrow ee)=30\%$,$BR (H^{--}\rightarrow \mu\mu)=40\%$ and $BR (H^{--}\rightarrow \mu e)=30\%$. We overlaid the current and projected ATLAS limits with 13TeV center of mass energy with $36.1fb^{-1}$ and $139fb^{-1}$ of integrated luminosity.}
    \label{fig:my_label1}
\end{figure}

The usual isolation criteria for dielectron and dimuon events take place \cite{Patra:2015bga,Lindner:2016lpp} to suppress the background coming from misidentified electrons as jets, which is prominent in the high $p_T$ regime. Moreover, jets within a cone of size $\sqrt{(\Delta \eta)^2 + (\Delta\phi)^2}=0.2$ are rejected, where  $\phi$ is the azimuthal angle around the beam line. It is known that the use of the inner tracking detector and muon spectrometer help to reduce the muon charge misidentification to a negligible level, thus the muon misidentification is not an issue in this type of study. The electron charge misidentification rate is more challenging though. 

Electron charge misidentification mostly results from bremsstrahlung. The emitted photon
might convert into an electron-positron pair toughening the charge reconstruction of the primary electron. Moreover, the emitted photon may, in some cases, travel through the inner detector without leaving a track. In the latter case, the electron charge, which is determined from its curvature track, is harder to determine especially for high-energy electrons which approximately travel straight, i.e making the measurement of the curvature more difficult. In the end, the charge misidentification rate induces roughly a 10\% systematic error in the analysis for electrons with $p_T > 100$~GeV \cite{ATLAS:2014kca}. 
The main backgrounds in this search are classified into three classes: prompt, misidentified leptons and non-prompt. The prompt background arises from SM processes producing two same-sign
leptons from $ZZ$, $WZ$, and $W^{\pm} W^{\pm}$ production, for instance. Opposite-sign lepton pairs produced from $W^{\pm} W^{\mp}$ also produce important background when the charge of one of the leptons is misidentified as aforementioned. The non-prompt background refers to jets misidentified as electrons and semileptonic decays of hadrons. In summary, one can clearly notice that the background sources do depend on the final state of interest.  

This is important to have in mind because the doubly charged Higgs decay width is found to be, 
\begin{equation}
    \Gamma (H^{\pm \pm}\rightarrow l^{\pm}l^{'\pm})= \frac{k\, h_{l l'}}{16 \pi} m_{H^{\pm \pm}},
\end{equation}where $k=1$ for $l\neq l'$ and $k=2$ for $l= l'$. 

Therefore, the couplings $h_{l l'}$ dictate the branching ratio into charged leptons and consequently the relevant final states. $h_{l l'}$ is assumed to be large enough (greater than $10^{-6}$) to ensure prompt decays. It is known that if this coupling is very small, displaced vertex searches become very important \cite{Han:2007bk,Dev:2018kpa,Antusch:2018svb}.

We emphasize that the decay into gauge bosons is assumed to be absent. In a more complete approach, this can be easily justified by taking the vacuum expectation value of the scalar triplet that gives rise to the doubly charged scalar to be sufficiently small. Moreover, in our analysis we will consider only decays into electron and muon final states motivated by the ATLAS and CMS collaborations approach. This assumption is very simplistic and deviations from it might induce a sizeable difference in the overall lower mass bounds. As highlighted earlier, the focus of this work is to project CMS and ATLAS sensitivity to such doubly charged scalars keeping the same assumption adopted by the collaborations.

That said, the relevant channel in our reasoning is the resonant pair production of doubly charged scalars as displayed in Fig.\ref{diagram1}. As we are carrying out our study in a model independent way, we will assume similarly to ATLAS and CMS collaboration that the decay width of the doubly charged scalar is small compared to the detector resolution. In other words, we will be working in the narrow width approximation. Besides the channel shown in Fig.\ref{diagram1} one could also have pair production of doubly charged scalars with photon-photon scattering where a doubly charged scalar would appear in the t-channel. However, in this case, the width of the doubly charged scalar becomes important and model dependent. We refer the reader to \cite{Babu:2016rcr,Crivellin:2018ahj} where the assessment of this channel was carefully investigated. In the end, the inclusion of this production channel improves the overall sensitivity to doubly charged scalars. In our work, we are mainly focused on the High-Luminosity and High-Energy LHC sensitivity to doubly charged scalars without considering a particular model and therefore we will not include these effects here. 

Concerning the pair production of doubly charged scalars decaying into $WW$ pairs which is present in the higgs triplet mode, this channel is suppressed by $v_{H}/M_W^2$, where $v_H$ is the vacuum expectation value of the scalar triplet. For sufficiently small values of the $v_H$ this channels is not relevant. There is another source of doubly charged scalar production which happens via its associated production with  $H^-$, $q\bar{q'} \rightarrow H^{++}H^{-}$. This channels it is not particularly clean since some decays go into neutrinos, missing energy. In summary, for these reasons we will focus on the pair production of $H^{++}H^{--}$ since it gives rise to a clean same sign dilepton signal.

Having discussed the signal and the simplifying assumptions adopted in our study we present the bounds in the next section.

\begin{figure}
    \includegraphics[scale=0.21]{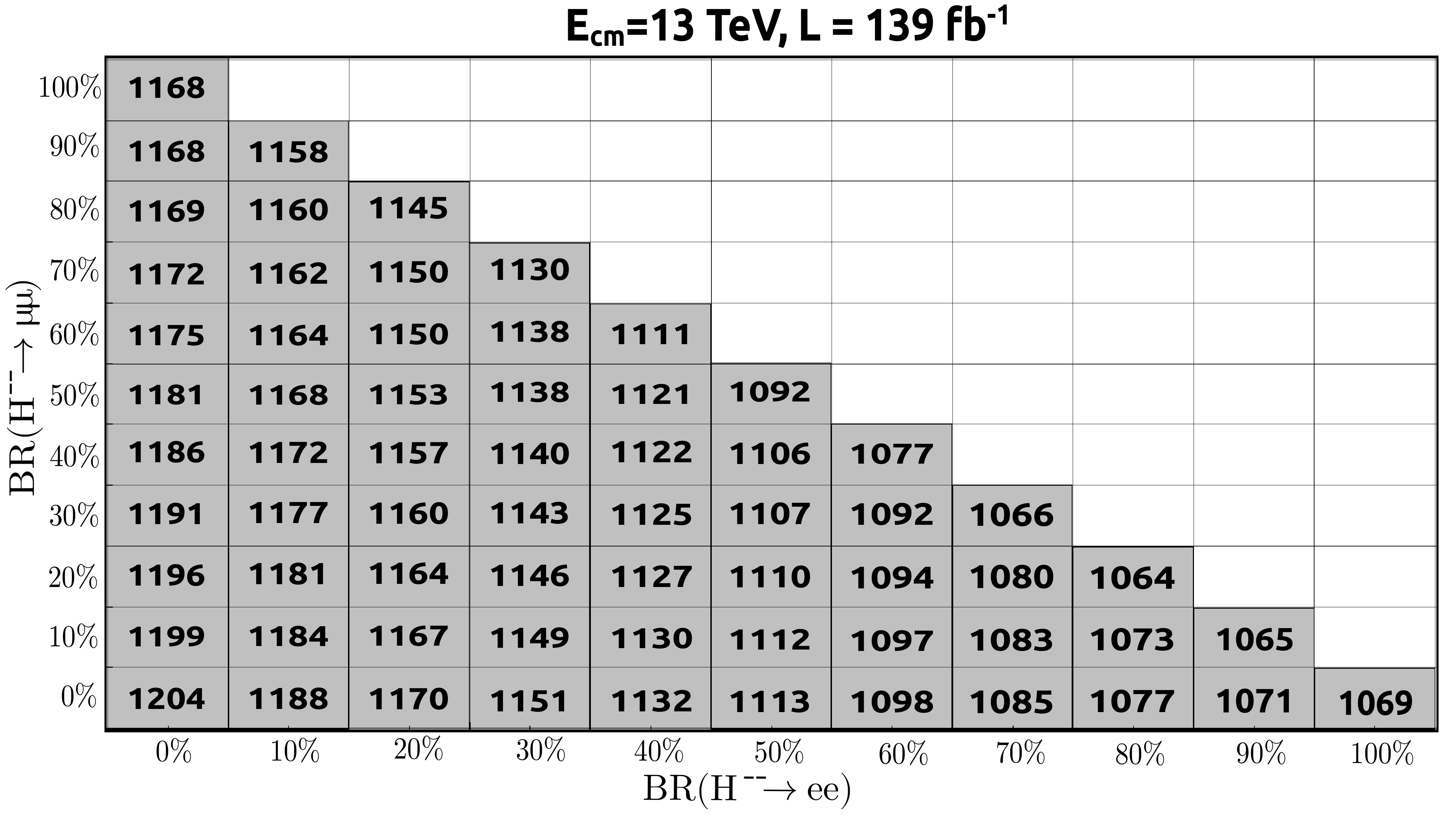}
    \caption{95\% C.L. lower mass bounds on the doubly charged scalar assuming BR $(H^{\pm\pm}\rightarrow l^{\pm}l^{\pm})=100\%$) for center of mass energy of $13$~TeV and $139\,fb^{-1}$ of integrated luminosity.}
    \label{fig:my_label3}
\end{figure}

\begin{figure}
    \includegraphics[scale=0.21]{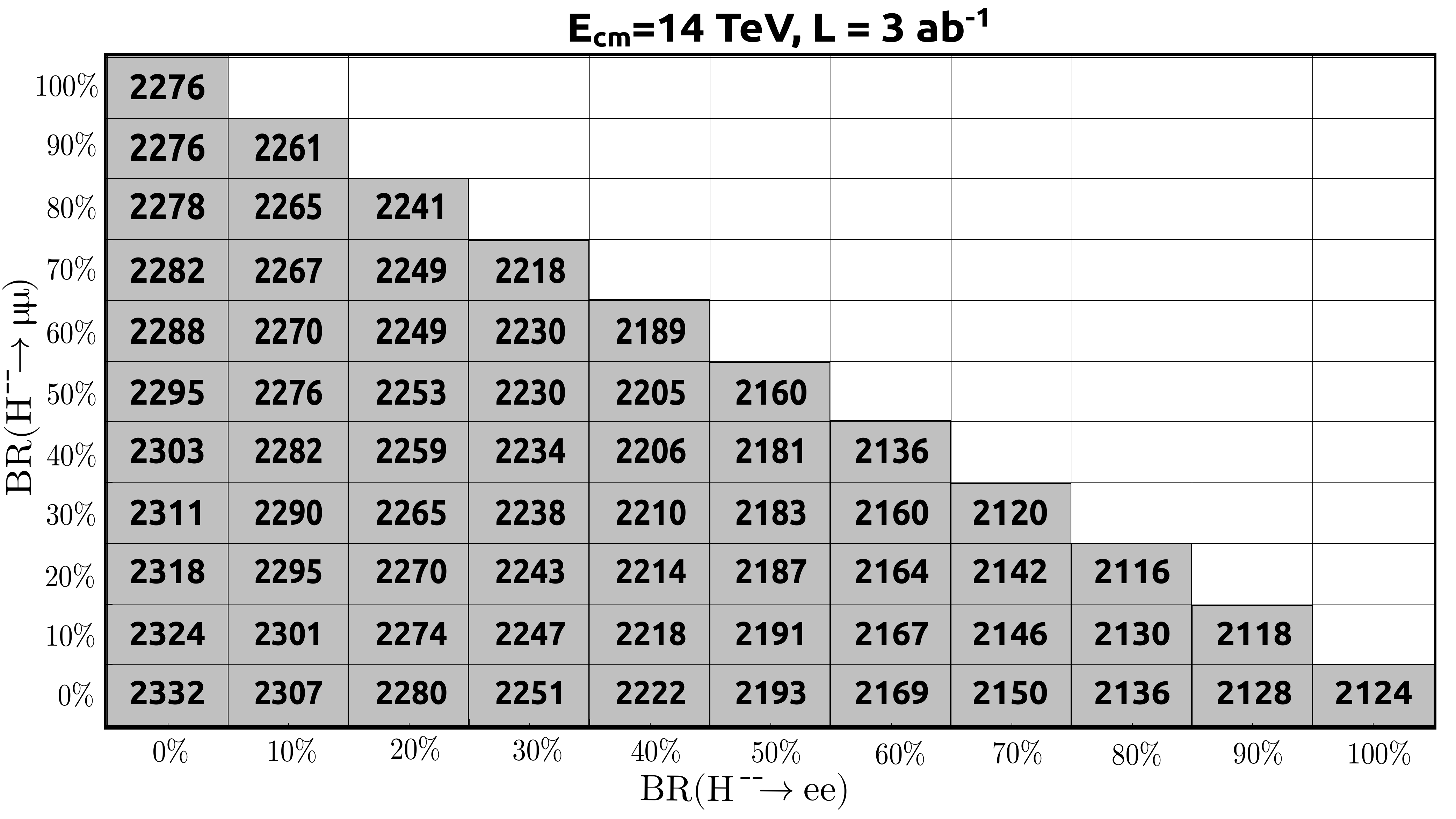}
    \caption{95\% C.L. lower mass bounds on the doubly charged scalar assuming BR $(H^{\pm\pm}\rightarrow l^{\pm}l^{\pm})=100\%$) for center of mass energy of $14$~TeV and $3\,ab^{-1}$ of integrated luminosity.}
    \label{fig:my_label4}
\end{figure}

\begin{figure}
    \includegraphics[scale=0.211]{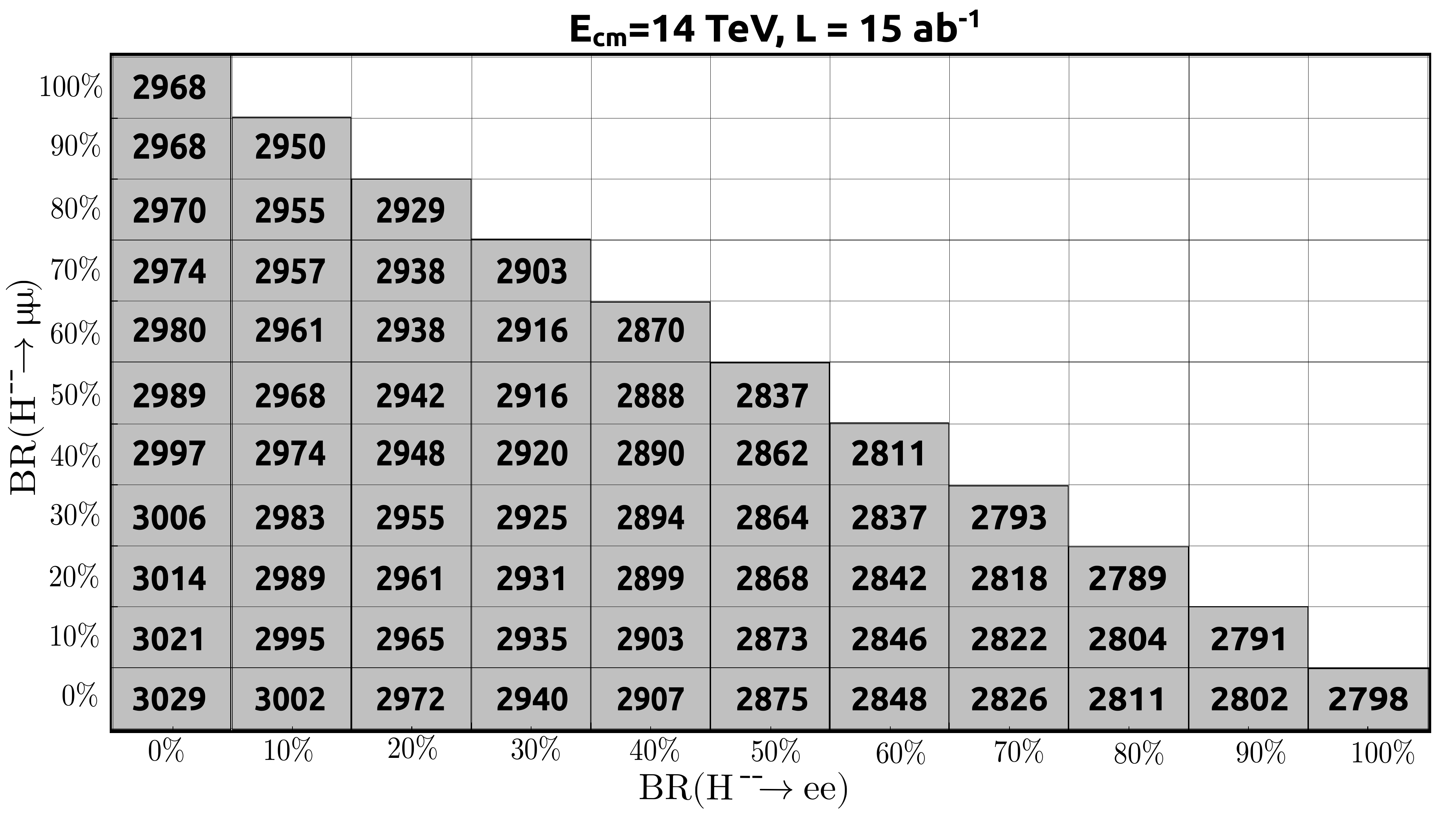}
    \caption{95\% C.L. lower mass bounds on the doubly charged scalar assuming BR $(H^{\pm\pm}\rightarrow l^{\pm}l^{\pm})=100\%$) for center of mass energy of $14$~TeV and  $15\,ab^{-1}$ of integrated luminosity.}
    \label{fig:my_label5}
\end{figure}

\begin{figure}
    \includegraphics[scale=0.211]{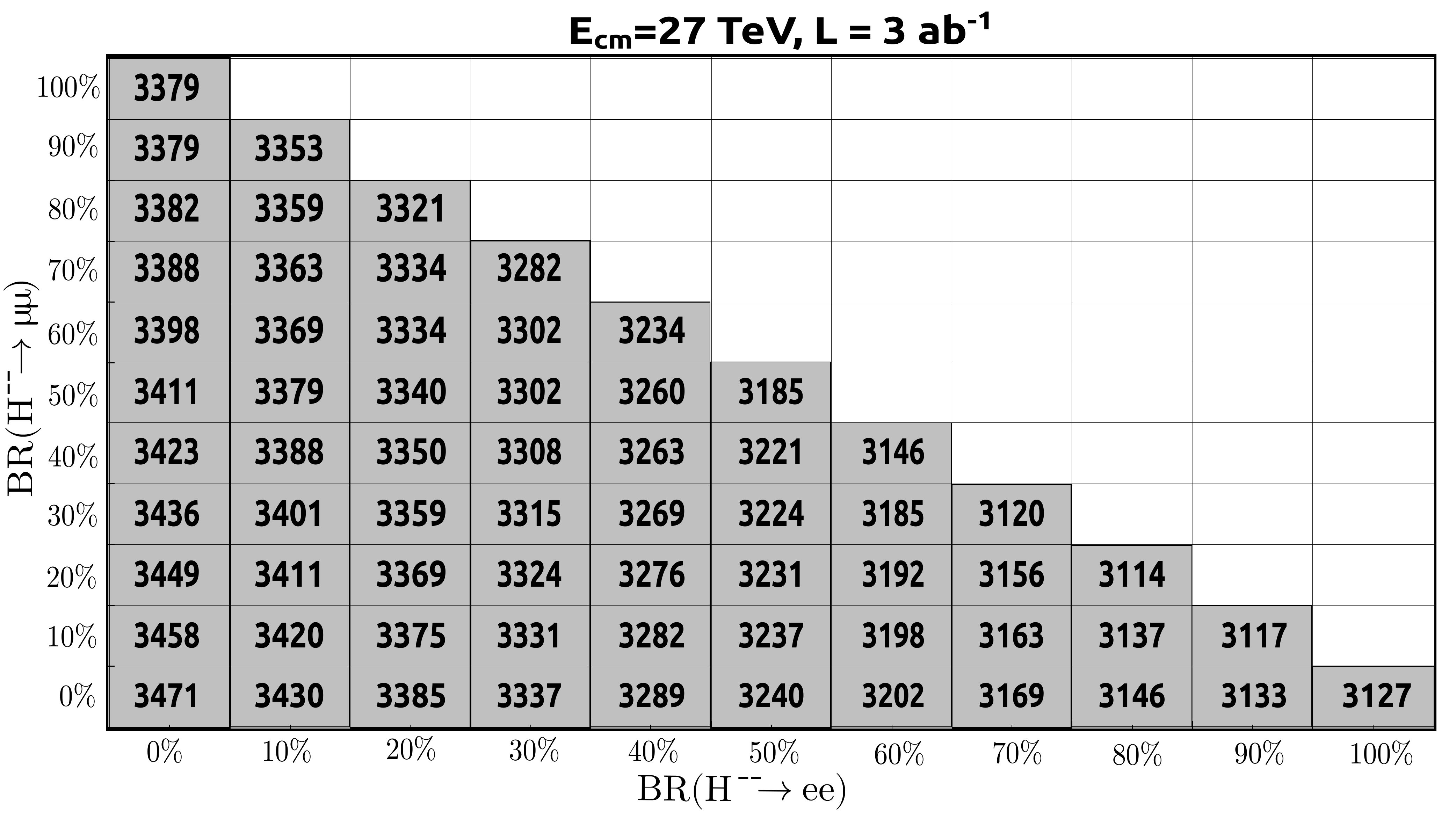}
    \caption{95\% C.L. lower mass bounds on the doubly charged scalar assuming BR $(H^{\pm\pm}\rightarrow l^{\pm}l^{\pm})=100\%$) for center of mass energy of $27$~TeV and $3\,ab^{-1}$ of integrated luminosity.}
    \label{fig:my_label6}
\end{figure}

\begin{figure}
    \includegraphics[scale=0.211]{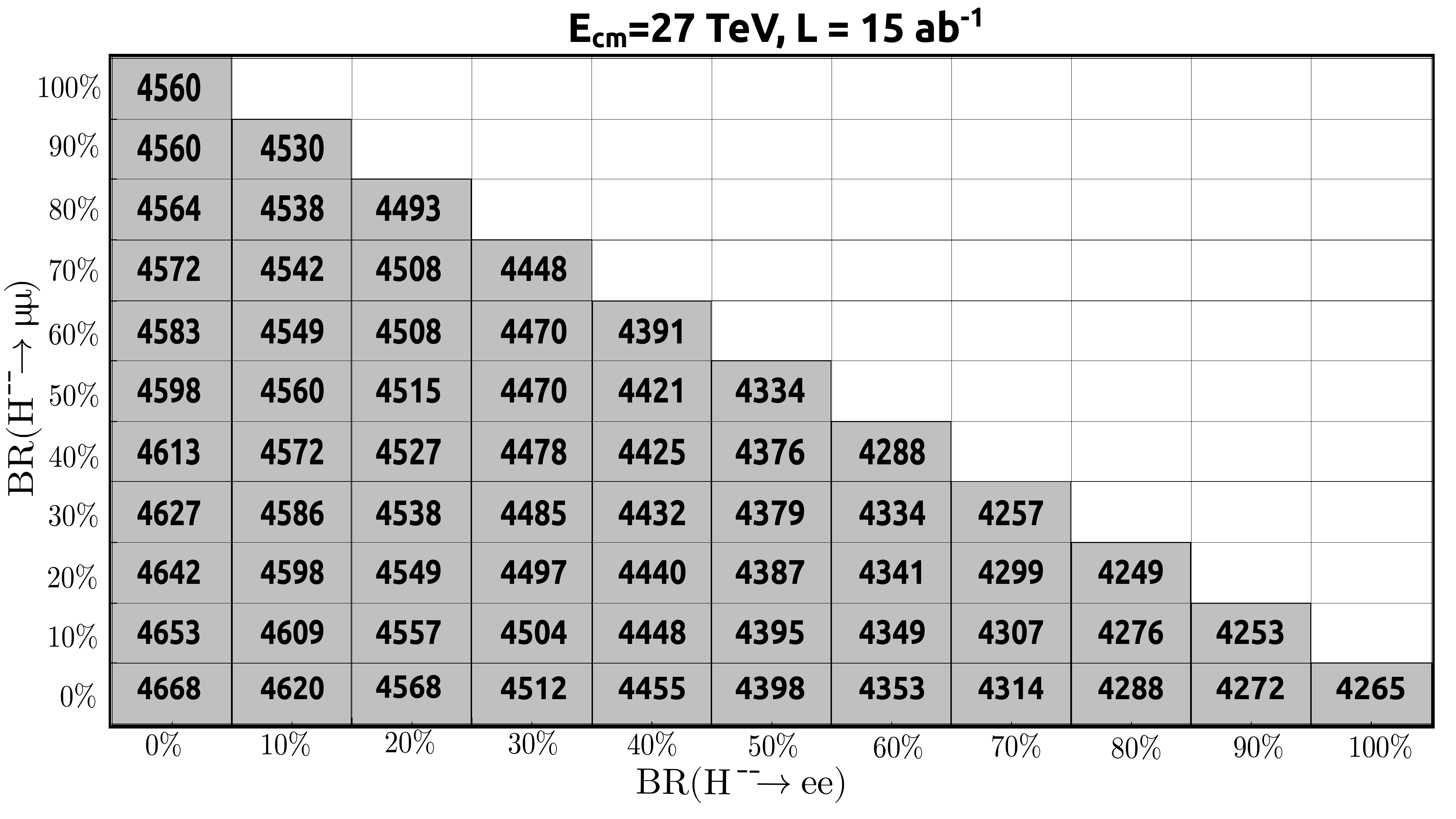}
    \caption{95\% C.L. lower mass bounds on the doubly charged scalar assuming BR $(H^{\pm\pm}\rightarrow l^{\pm}l^{\pm})=100\%$) for center of mass energy of $27$~TeV and $15\, ab^{-1}$ of integrated luminosity.}
    \label{fig:my_label7}
\end{figure}

\section{Bounds}

ATLAS performed a resonant production search for doubly charged scalars, based on the diagram shown in Fig.\ref{diagram1} assuming $BR (H^{\pm \pm} \rightarrow e^{\pm}e^{\pm})+BR (H^{\pm \pm} \rightarrow e^{\pm}\mu^{\pm})+BR (H^{\pm \pm} \rightarrow \mu^{\pm}\mu^{\pm})=100\%$ taking into account all the relevant background discussed earlier and reported null results in agreement with CMS collaboration \cite{CMS:2017pet}. 
As a result, an upper limit on the production cross section was derived for different decay modes. In Fig.\ref{fig:my_label1} we exhibit the 95\% C.L. limit on the production cross section derived by ATLAS using $36.1 fb^{-1}$ of data based on $13$~TeV proton-proton collisions.  We computed the production cross section of the doubly charged scalar using MadGraph5 \cite{Alwall:2011uj} and  included hadronization and detector effects via Pythia \cite{Sjostrand:2007gs} and Delphes \cite{deFavereau:2013fsa} interfaces considering all the cuts described previously to impose $m_{H^{\pm\pm}} > 840$~GeV, as one can see in Fig.\ref{fig:my_label1}. Our result is in excellent agreement with ATLAS collaboration.

In order to forecast lower mass bounds for new collider configurations, one would need to scale the backgrounds and then find a new point that yields the same number of background events, assuming the same detector efficiency and acceptance. As we are dealing with dilepton events, the number of signal and background events scale equally with energy and luminosity allowing us to forecast the future collider sensitivities. In other words, the upper limit obtained on the number of signal events at each mass point is derived as a function of the number of background events. This is valid if the signal acceptance and efficiency are nearly independent of the resonance mass and center of mass energy. We emphasize that our assumptions are only reasonable for resonance searches, which do not need to rely on shape analysis. For this reason, we can use the code described in \cite{Thamm:2015zwa} to obtain future proton-proton collider sensitivities. In simple terms, we need to solve an equation of the type $N(m_{H^{\pm\pm}}^2,E_{new},\mathcal{L}_{new})/N({\rm bound},13TeV,36.1 fb^{1})=1$, where $N$ is the number of signal events for a given parton distribution function which we assumed to be the NNPDF23-NNLO \cite{Ball:2014uwa}.

Firstly, we carried out this procedure to obtain the LHC sensitivity with $139 fb^{-1}$ of integrated luminosity, which should be out in the next public LHC data release. We assumed that the doubly charged scalar decays $30\%$ into $ee$, $40\%$ into $\mu\mu$ and $30\%$ into $e\mu$ and derived the lower mass bound as exhibited in Fig.\ref{fig:my_label1}. We repeated this exercise for a multitude of possibilities varying the branching into $ee$, $\mu\mu$ and $e\mu$ as shown in Fig.\ref{fig:my_label3}.

Looking at Fig.\ref{fig:my_label3} one can conclude that the larger the branching ratio into $ee$ the weaker the bound. This has to do partially with the effects we described earlier such as the large electron misidentification rate which has shown to be important in other collider sensitivity studies \cite{Lindner:2016lxq,Queiroz:2016qmc}. In particular, in case of null results one might impose $m_{H^{\pm\pm}}>1.2$~TeV assuming the doubly charged scalar decaying entirely into $e\mu$ pairs, i.e. with $BR(H^{--}\rightarrow ee)=0$ and $BR(H^{--}\rightarrow \mu\mu)=0$. Although, notice that if we keep $BR(H^{--}\rightarrow ee)=0$ and increase the branching ratio into muons the limits weakens. Therefore, our simple interpretation of the observed limits is not entirely related to the electron misidentification rate. If one cares to compare the expected limits and the observed limits obtained by ATLAS collaboration \cite{Aaboud:2017qph} one will clearly conclude that electron misidentification is the driving force behind the weakening of the bounds as branching ratio into $ee$ increases. If we keep $BR(H^{--}\rightarrow ee)=0\%$ and start from $BR(H^{--}\rightarrow \mu\mu)=100\%$ we can notice that as we decrease the branching ratio into $\mu\mu$ the bound strenthens and this is because we are inscreasing the branching ratio into the lepton flavor violation channel which suffers from a relatively small Standard Model background. Therefore, the combinantion of electron misidentification rate with the decay into lepton flavor violating decays helps us understand the lower mass bounds in the figures.

Now considering the High-Luminosity LHC (HL-LHC) setup, which is based on $14$~TeV center of mass energy and $\mathcal{L}=3 ab^{-1}$ we derived Fig.\ref{fig:my_label4}. The HL-LHC can probe masses up to $2.3$~TeV going down to $2.1$~TeV when the branching ratio into $ee$ is around $100\%$. We also derived the HL-LHC sensitivity when the luminosity is ramped up to $\mathcal{L}=15 ab^{-1}$ exhibited in Fig.\ref{fig:my_label5}. The  High-Energy LHC (HE-LHC) reach was chosen to be the $27$~TeV center of mass energy proton-proton collision with  $\mathcal{L}=15 ab^{-1}$, according to the report \cite{CidVidal:2018eel}, but for completeness we also found the HE-LHC reach for $\mathcal{L}=3 ab^{-1}$. 

Looking at Fig.\ref{fig:my_label6}-\ref{fig:my_label7} we see that HE-LHC will be able to explore doubly charged scalars with masses of $3.1$ TeV with $\mathcal{L}=3 ab^{-1}$, all the way up to $4.6$~TeV with $\mathcal{L}=3 ab^{-1}$. These projected bounds certainly constitute the strongest ones on the doubly charged scalar mass surpassing those coming from lepton flavor violation searches \cite{Lindner:2016bgg,Dev:2018sel,Ferreira:2019qpf}, particularly in the case where the coupling constant involving $e\mu$ is negligible.

We highlight that if doubly charged scalar has sizeable couplings to $e\mu$, searches for lepton flavor violating muon decays lead to stronger bounds on the doubly charged scalar mass \cite{Lindner:2016bgg}. Collider bounds are more relevant otherwise.

Having in mind that other future collider projections aim at probing doubly charged scalars up to $2$~TeV \cite{Dev:2019hev} we conclude that in the absence of lepton flavor violation decays HE-LHC will give rise to the most stringent lower mass bounds in case of null results. 

\section{Photon-Photon Fusion}

In this section we will present our forecast for the HL-LHC and HE-LHC for the case where photon-photon fusion processes are included in the analysis. In this case there are many diagrams that contribute. Adding them we inscrease the production cross section up to roughly 60\% \cite{Babu:2016rcr}. We will keep assuming that only ($e^{\pm} e^{\pm},\mu^{\pm} \mu^{\pm}$, and $e^{\pm}\mu^{\pm}$) might appear in the final states, though. This time we will focus on a few benchmark cases. Our lower mass bounds when the Drell-Yann and photon initiated processes are  summarized in the table I. 

\begin{table}
\begin{tabular}{|c|c|c|}
\hline
Scenario & HL-LHC & HE-LHC\\
BR$(H^{--}\rightarrow ee)=100\%$ & 3.13 TeV & 4.84 TeV\\
\hline
BR$(H^{--}\rightarrow e\mu)=100\%$ & 3 TeV & 4.72 TeV\\
\hline
BR$(H^{--}\rightarrow \mu\mu)=100\%$ & 3.1 TeV & 4.79 TeV\\
\hline
\end{tabular}
\caption{The HL-LHC and HE-LHC sensitivity to doubly charged scalar when the Drell-Yann and photon-photon fusion processes are included. HL-LHC and HE-LHC refer to the LHC running at $14$~TeV and $27$~TeV with $\mathcal{L}=15 ab^{-1}$, respectively.}
\end{table}

One can easily conclude tha the inclusion of photon-photon mediated processes significantly improve the overall lower mass bounds. For instance, in the HE-LHC setup for $BR(H^{--}\rightarrow ee)=100\%$ the limits went from $4.265$~TeV to $4.84$~TeV. One should keep in mind that at those energies the proton PDF brings a significant uncertainty to the photon-photon mediated fusion processes, thus this bounds should be taken with care. 

\section{Comparison with future  Colliders}

In this section we compare our projected lower mass limits with the existing ones stemming from future colliders \cite{Dev:2018upe}. At the International Linear Collider, there are two processes that lead to stringent collider bounds on the doubly charged scalar. The on-shell pair production of doubly charged scalars will take place for $m_{H^{\pm\pm}} <500$~GeV, so one can automatically conclude that HL-LHC and HE-LHC constitues a better probe, leaving to ILC the role for more precise measurements \cite{Nomura:2017abh}. The off-shell production occurs via  u-channel exchange of the doubly charged scalar \cite{Swartz:1989qz}. In this case the constraints on the doubly charge scalar mass depend very much in the yukawa couplings. Anyway, the overall ILC sensitivity is not comparable to the HE-LHC. In other words, HL-LHC and HE-LHC still represent more promissing probes for such scalars at the TeV scale \cite{Nomura:2017abh}. Regarding the 100TeV collider, the situation changes for obvious reasons. The great improvement in the center-of-mass energy allows this collider to probe doubly charged scalars up to $10$~TeV, and even claim discovery for masses below $4.5$~TeV \cite{Du:2018eaw}, something not feasible within the HE-LHC.

\section{Conclusions}

Doubly charged scalars are present in several new physics models such as Higgs triplets, type II seesaw mechanism, left-right and 3-3-1 models. They play an important role in neutrino masses and lepton flavor violation observables. Motivated by their overall importance we obtained collider bounds under reasonable decay assumptions, similarly adopted by ATLAS collaboration. We started forecasting the LHC reach to doubly charged scalars for an integrated luminosity of $\mathcal{L}=139fb^{-1}$ varying the branching ratio into charged leptons. Moreover, we obtained the High-Luminosity and High-Energy LHC sensitivity to doubly charged scalars running with $3 ab^{-1}$ and $15ab^{-1}$ of integrated luminosity. As we varied the branching ratio into charged leptons for each collider configuration studied, in the end, we ended up covering fifty possible decay channels and consequently deriving fifty lower mass bounds in each collider setup considered. In particular, we showed that the HE-LHC will be able to probe doubly charged scalar masses up to $\sim 4.33$~TeV when it decays exclusively into ee and $\mu\mu$ final states, and masses of $4.66$~TeV when it decays entirely into $e\mu$. The latter case of flavor changing decays is complementary to the one stemming from lepton flavor violation. Anyway, it is clear that the HE-LHC will constitute a powerful probe for doubly charged scalars. 

\section*{Acknowledgements}
The authors thank Diego Restrepo and Clarissa Siqueira for discussions. FSQ acknowledges support from CNPq
grants 303817/2018-6 and 421952/2018-0, UFRN, MEC and
ICTP-SAIFR FAPESP grant 2016/01343-7. TM and YV acknowledge CAPES fellowship.  
%\clearpage

\bibliography{ref}

%merlin.mbs apsrev4-1.bst 2010-07-25 4.21a (PWD, AO, DPC) hacked
%Control: key (0)
%Control: author (8) initials jnrlst
%Control: editor formatted (1) identically to author
%Control: production of article title (-1) disabled
%Control: page (0) single
%Control: year (1) truncated
%Control: production of eprint (0) enabled
\begin{thebibliography}{77}%
\makeatletter
\providecommand \@ifxundefined [1]{%
 \@ifx{#1\undefined}
}%
\providecommand \@ifnum [1]{%
 \ifnum #1\expandafter \@firstoftwo
 \else \expandafter \@secondoftwo
 \fi
}%
\providecommand \@ifx [1]{%
 \ifx #1\expandafter \@firstoftwo
 \else \expandafter \@secondoftwo
 \fi
}%
\providecommand \natexlab [1]{#1}%
\providecommand \enquote  [1]{``#1''}%
\providecommand \bibnamefont  [1]{#1}%
\providecommand \bibfnamefont [1]{#1}%
\providecommand \citenamefont [1]{#1}%
\providecommand \href@noop [0]{\@secondoftwo}%
\providecommand \href [0]{\begingroup \@sanitize@url \@href}%
\providecommand \@href[1]{\@@startlink{#1}\@@href}%
\providecommand \@@href[1]{\endgroup#1\@@endlink}%
\providecommand \@sanitize@url [0]{\catcode `\\12\catcode `\$12\catcode
  `\&12\catcode `\#12\catcode `\^12\catcode `\_12\catcode `\%12\relax}%
\providecommand \@@startlink[1]{}%
\providecommand \@@endlink[0]{}%
\providecommand \url  [0]{\begingroup\@sanitize@url \@url }%
\providecommand \@url [1]{\endgroup\@href {#1}{\urlprefix }}%
\providecommand \urlprefix  [0]{URL }%
\providecommand \Eprint [0]{\href }%
\providecommand \doibase [0]{http://dx.doi.org/}%
\providecommand \selectlanguage [0]{\@gobble}%
\providecommand \bibinfo  [0]{\@secondoftwo}%
\providecommand \bibfield  [0]{\@secondoftwo}%
\providecommand \translation [1]{[#1]}%
\providecommand \BibitemOpen [0]{}%
\providecommand \bibitemStop [0]{}%
\providecommand \bibitemNoStop [0]{.\EOS\space}%
\providecommand \EOS [0]{\spacefactor3000\relax}%
\providecommand \BibitemShut  [1]{\csname bibitem#1\endcsname}%
\let\auto@bib@innerbib\@empty
%</preamble>
\bibitem [{\citenamefont {Muhlleitner}\ and\ \citenamefont
  {Spira}(2003)}]{Muhlleitner:2003me}%
  \BibitemOpen
  \bibfield  {author} {\bibinfo {author} {\bibfnamefont {M.}~\bibnamefont
  {Muhlleitner}}\ and\ \bibinfo {author} {\bibfnamefont {M.}~\bibnamefont
  {Spira}},\ }\href {\doibase 10.1103/PhysRevD.68.117701} {\bibfield  {journal}
  {\bibinfo  {journal} {Phys. Rev.}\ }\textbf {\bibinfo {volume} {D68}},\
  \bibinfo {pages} {117701} (\bibinfo {year} {2003})},\ \Eprint
  {http://arxiv.org/abs/hep-ph/0305288} {arXiv:hep-ph/0305288 [hep-ph]}
  \BibitemShut {NoStop}%
%%CITATION = HEP-PH/0305288;%%
\bibitem [{\citenamefont {Akeroyd}\ and\ \citenamefont
  {Aoki}(2005)}]{Akeroyd:2005gt}%
  \BibitemOpen
  \bibfield  {author} {\bibinfo {author} {\bibfnamefont {A.~G.}\ \bibnamefont
  {Akeroyd}}\ and\ \bibinfo {author} {\bibfnamefont {M.}~\bibnamefont {Aoki}},\
  }\href {\doibase 10.1103/PhysRevD.72.035011} {\bibfield  {journal} {\bibinfo
  {journal} {Phys. Rev.}\ }\textbf {\bibinfo {volume} {D72}},\ \bibinfo {pages}
  {035011} (\bibinfo {year} {2005})},\ \Eprint
  {http://arxiv.org/abs/hep-ph/0506176} {arXiv:hep-ph/0506176 [hep-ph]}
  \BibitemShut {NoStop}%
%%CITATION = HEP-PH/0506176;%%
\bibitem [{\citenamefont {Hektor}\ \emph {et~al.}(2007)\citenamefont {Hektor},
  \citenamefont {Kadastik}, \citenamefont {Muntel}, \citenamefont {Raidal},\
  and\ \citenamefont {Rebane}}]{Hektor:2007uu}%
  \BibitemOpen
  \bibfield  {author} {\bibinfo {author} {\bibfnamefont {A.}~\bibnamefont
  {Hektor}}, \bibinfo {author} {\bibfnamefont {M.}~\bibnamefont {Kadastik}},
  \bibinfo {author} {\bibfnamefont {M.}~\bibnamefont {Muntel}}, \bibinfo
  {author} {\bibfnamefont {M.}~\bibnamefont {Raidal}}, \ and\ \bibinfo {author}
  {\bibfnamefont {L.}~\bibnamefont {Rebane}},\ }\href {\doibase
  10.1016/j.nuclphysb.2007.07.014} {\bibfield  {journal} {\bibinfo  {journal}
  {Nucl. Phys.}\ }\textbf {\bibinfo {volume} {B787}},\ \bibinfo {pages} {198}
  (\bibinfo {year} {2007})},\ \Eprint {http://arxiv.org/abs/0705.1495}
  {arXiv:0705.1495 [hep-ph]} \BibitemShut {NoStop}%
%%CITATION = ARXIV:0705.1495;%%
\bibitem [{\citenamefont {Fileviez~Perez}\ \emph
  {et~al.}(2008{\natexlab{a}})\citenamefont {Fileviez~Perez}, \citenamefont
  {Han}, \citenamefont {Huang}, \citenamefont {Li},\ and\ \citenamefont
  {Wang}}]{Perez:2008zc}%
  \BibitemOpen
  \bibfield  {author} {\bibinfo {author} {\bibfnamefont {P.}~\bibnamefont
  {Fileviez~Perez}}, \bibinfo {author} {\bibfnamefont {T.}~\bibnamefont {Han}},
  \bibinfo {author} {\bibfnamefont {G.-Y.}\ \bibnamefont {Huang}}, \bibinfo
  {author} {\bibfnamefont {T.}~\bibnamefont {Li}}, \ and\ \bibinfo {author}
  {\bibfnamefont {K.}~\bibnamefont {Wang}},\ }\href {\doibase
  10.1103/PhysRevD.78.071301} {\bibfield  {journal} {\bibinfo  {journal} {Phys.
  Rev.}\ }\textbf {\bibinfo {volume} {D78}},\ \bibinfo {pages} {071301}
  (\bibinfo {year} {2008}{\natexlab{a}})},\ \Eprint
  {http://arxiv.org/abs/0803.3450} {arXiv:0803.3450 [hep-ph]} \BibitemShut
  {NoStop}%
%%CITATION = ARXIV:0803.3450;%%
\bibitem [{\citenamefont {Chaudhuri}\ \emph {et~al.}(2014)\citenamefont
  {Chaudhuri}, \citenamefont {Grimus},\ and\ \citenamefont
  {Mukhopadhyaya}}]{Chaudhuri:2013xoa}%
  \BibitemOpen
  \bibfield  {author} {\bibinfo {author} {\bibfnamefont {A.}~\bibnamefont
  {Chaudhuri}}, \bibinfo {author} {\bibfnamefont {W.}~\bibnamefont {Grimus}}, \
  and\ \bibinfo {author} {\bibfnamefont {B.}~\bibnamefont {Mukhopadhyaya}},\
  }\href {\doibase 10.1007/JHEP02(2014)060} {\bibfield  {journal} {\bibinfo
  {journal} {JHEP}\ }\textbf {\bibinfo {volume} {02}},\ \bibinfo {pages} {060}
  (\bibinfo {year} {2014})},\ \Eprint {http://arxiv.org/abs/1305.5761}
  {arXiv:1305.5761 [hep-ph]} \BibitemShut {NoStop}%
%%CITATION = ARXIV:1305.5761;%%
\bibitem [{\citenamefont {Lindner}\ \emph {et~al.}(2018)\citenamefont
  {Lindner}, \citenamefont {Platscher},\ and\ \citenamefont
  {Queiroz}}]{Lindner:2016bgg}%
  \BibitemOpen
  \bibfield  {author} {\bibinfo {author} {\bibfnamefont {M.}~\bibnamefont
  {Lindner}}, \bibinfo {author} {\bibfnamefont {M.}~\bibnamefont {Platscher}},
  \ and\ \bibinfo {author} {\bibfnamefont {F.~S.}\ \bibnamefont {Queiroz}},\
  }\href {\doibase 10.1016/j.physrep.2017.12.001} {\bibfield  {journal}
  {\bibinfo  {journal} {Phys. Rept.}\ }\textbf {\bibinfo {volume} {731}},\
  \bibinfo {pages} {1} (\bibinfo {year} {2018})},\ \Eprint
  {http://arxiv.org/abs/1610.06587} {arXiv:1610.06587 [hep-ph]} \BibitemShut
  {NoStop}%
%%CITATION = ARXIV:1610.06587;%%
\bibitem [{\citenamefont {Primulando}\ \emph {et~al.}(2019)\citenamefont
  {Primulando}, \citenamefont {Julio},\ and\ \citenamefont
  {Uttayarat}}]{Primulando:2019evb}%
  \BibitemOpen
  \bibfield  {author} {\bibinfo {author} {\bibfnamefont {R.}~\bibnamefont
  {Primulando}}, \bibinfo {author} {\bibfnamefont {J.}~\bibnamefont {Julio}}, \
  and\ \bibinfo {author} {\bibfnamefont {P.}~\bibnamefont {Uttayarat}},\
  }\href@noop {} {\  (\bibinfo {year} {2019})},\ \Eprint
  {http://arxiv.org/abs/1903.02493} {arXiv:1903.02493 [hep-ph]} \BibitemShut
  {NoStop}%
%%CITATION = ARXIV:1903.02493;%%
\bibitem [{\citenamefont {Pati}\ and\ \citenamefont
  {Salam}(1974)}]{Pati:1974yy}%
  \BibitemOpen
  \bibfield  {author} {\bibinfo {author} {\bibfnamefont {J.~C.}\ \bibnamefont
  {Pati}}\ and\ \bibinfo {author} {\bibfnamefont {A.}~\bibnamefont {Salam}},\
  }\href {\doibase 10.1103/PhysRevD.10.275, 10.1103/PhysRevD.11.703.2}
  {\bibfield  {journal} {\bibinfo  {journal} {Phys. Rev.}\ }\textbf {\bibinfo
  {volume} {D10}},\ \bibinfo {pages} {275} (\bibinfo {year} {1974})},\ \bibinfo
  {note} {[Erratum: Phys. Rev.D11,703(1975)]}\BibitemShut {NoStop}%
%%CITATION = PHRVA,D10,275;%%
\bibitem [{\citenamefont {Mohapatra}\ and\ \citenamefont
  {Pati}(1975)}]{Mohapatra:1974hk}%
  \BibitemOpen
  \bibfield  {author} {\bibinfo {author} {\bibfnamefont {R.~N.}\ \bibnamefont
  {Mohapatra}}\ and\ \bibinfo {author} {\bibfnamefont {J.~C.}\ \bibnamefont
  {Pati}},\ }\href {\doibase 10.1103/PhysRevD.11.566} {\bibfield  {journal}
  {\bibinfo  {journal} {Phys. Rev.}\ }\textbf {\bibinfo {volume} {D11}},\
  \bibinfo {pages} {566} (\bibinfo {year} {1975})}\BibitemShut {NoStop}%
%%CITATION = PHRVA,D11,566;%%
\bibitem [{\citenamefont {Senjanovic}\ and\ \citenamefont
  {Mohapatra}(1975)}]{Senjanovic:1975rk}%
  \BibitemOpen
  \bibfield  {author} {\bibinfo {author} {\bibfnamefont {G.}~\bibnamefont
  {Senjanovic}}\ and\ \bibinfo {author} {\bibfnamefont {R.~N.}\ \bibnamefont
  {Mohapatra}},\ }\href {\doibase 10.1103/PhysRevD.12.1502} {\bibfield
  {journal} {\bibinfo  {journal} {Phys. Rev.}\ }\textbf {\bibinfo {volume}
  {D12}},\ \bibinfo {pages} {1502} (\bibinfo {year} {1975})}\BibitemShut
  {NoStop}%
%%CITATION = PHRVA,D12,1502;%%
\bibitem [{\citenamefont {Dutta}\ \emph {et~al.}(2014)\citenamefont {Dutta},
  \citenamefont {Eusebi}, \citenamefont {Gao}, \citenamefont {Ghosh},\ and\
  \citenamefont {Kamon}}]{Dutta:2014dba}%
  \BibitemOpen
  \bibfield  {author} {\bibinfo {author} {\bibfnamefont {B.}~\bibnamefont
  {Dutta}}, \bibinfo {author} {\bibfnamefont {R.}~\bibnamefont {Eusebi}},
  \bibinfo {author} {\bibfnamefont {Y.}~\bibnamefont {Gao}}, \bibinfo {author}
  {\bibfnamefont {T.}~\bibnamefont {Ghosh}}, \ and\ \bibinfo {author}
  {\bibfnamefont {T.}~\bibnamefont {Kamon}},\ }\href {\doibase
  10.1103/PhysRevD.90.055015} {\bibfield  {journal} {\bibinfo  {journal} {Phys.
  Rev.}\ }\textbf {\bibinfo {volume} {D90}},\ \bibinfo {pages} {055015}
  (\bibinfo {year} {2014})},\ \Eprint {http://arxiv.org/abs/1404.0685}
  {arXiv:1404.0685 [hep-ph]} \BibitemShut {NoStop}%
%%CITATION = ARXIV:1404.0685;%%
\bibitem [{\citenamefont {Bhupal~Dev}\ \emph {et~al.}(2017)\citenamefont
  {Bhupal~Dev}, \citenamefont {Mohapatra},\ and\ \citenamefont
  {Zhang}}]{Dev:2016vle}%
  \BibitemOpen
  \bibfield  {author} {\bibinfo {author} {\bibfnamefont {P.~S.}\ \bibnamefont
  {Bhupal~Dev}}, \bibinfo {author} {\bibfnamefont {R.~N.}\ \bibnamefont
  {Mohapatra}}, \ and\ \bibinfo {author} {\bibfnamefont {Y.}~\bibnamefont
  {Zhang}},\ }\href {\doibase 10.1103/PhysRevD.95.115001} {\bibfield  {journal}
  {\bibinfo  {journal} {Phys. Rev.}\ }\textbf {\bibinfo {volume} {D95}},\
  \bibinfo {pages} {115001} (\bibinfo {year} {2017})},\ \Eprint
  {http://arxiv.org/abs/1612.09587} {arXiv:1612.09587 [hep-ph]} \BibitemShut
  {NoStop}%
%%CITATION = ARXIV:1612.09587;%%
\bibitem [{\citenamefont {Borah}\ and\ \citenamefont
  {Dasgupta}(2017)}]{Borah:2016hqn}%
  \BibitemOpen
  \bibfield  {author} {\bibinfo {author} {\bibfnamefont {D.}~\bibnamefont
  {Borah}}\ and\ \bibinfo {author} {\bibfnamefont {A.}~\bibnamefont
  {Dasgupta}},\ }\href {\doibase 10.1007/JHEP01(2017)072} {\bibfield  {journal}
  {\bibinfo  {journal} {JHEP}\ }\textbf {\bibinfo {volume} {01}},\ \bibinfo
  {pages} {072} (\bibinfo {year} {2017})},\ \Eprint
  {http://arxiv.org/abs/1609.04236} {arXiv:1609.04236 [hep-ph]} \BibitemShut
  {NoStop}%
%%CITATION = ARXIV:1609.04236;%%
\bibitem [{\citenamefont {Bhupal~Dev}\ and\ \citenamefont
  {Zhang}(2018)}]{Dev:2018kpa}%
  \BibitemOpen
  \bibfield  {author} {\bibinfo {author} {\bibfnamefont {P.~S.}\ \bibnamefont
  {Bhupal~Dev}}\ and\ \bibinfo {author} {\bibfnamefont {Y.}~\bibnamefont
  {Zhang}},\ }\href {\doibase 10.1007/JHEP10(2018)199} {\bibfield  {journal}
  {\bibinfo  {journal} {JHEP}\ }\textbf {\bibinfo {volume} {10}},\ \bibinfo
  {pages} {199} (\bibinfo {year} {2018})},\ \Eprint
  {http://arxiv.org/abs/1808.00943} {arXiv:1808.00943 [hep-ph]} \BibitemShut
  {NoStop}%
%%CITATION = ARXIV:1808.00943;%%
\bibitem [{\citenamefont {Cieza~Montalvo}\ \emph {et~al.}(2006)\citenamefont
  {Cieza~Montalvo}, \citenamefont {Cortez}, \citenamefont {Sa~Borges},\ and\
  \citenamefont {Tonasse}}]{CiezaMontalvo:2006zt}%
  \BibitemOpen
  \bibfield  {author} {\bibinfo {author} {\bibfnamefont {J.~E.}\ \bibnamefont
  {Cieza~Montalvo}}, \bibinfo {author} {\bibfnamefont {N.~V.}\ \bibnamefont
  {Cortez}}, \bibinfo {author} {\bibfnamefont {J.}~\bibnamefont {Sa~Borges}}, \
  and\ \bibinfo {author} {\bibfnamefont {M.~D.}\ \bibnamefont {Tonasse}},\
  }\href {\doibase 10.1016/j.nuclphysb.2006.08.013,
  10.1016/j.nuclphysb.2008.01.003} {\bibfield  {journal} {\bibinfo  {journal}
  {Nucl. Phys.}\ }\textbf {\bibinfo {volume} {B756}},\ \bibinfo {pages} {1}
  (\bibinfo {year} {2006})},\ \bibinfo {note} {[Erratum: Nucl.
  Phys.B796,422(2008)]},\ \Eprint {http://arxiv.org/abs/hep-ph/0606243}
  {arXiv:hep-ph/0606243 [hep-ph]} \BibitemShut {NoStop}%
%%CITATION = HEP-PH/0606243;%%
\bibitem [{\citenamefont {Alves}\ \emph {et~al.}(2011)\citenamefont {Alves},
  \citenamefont {Ramirez~Barreto}, \citenamefont {Dias}, \citenamefont
  {de~S.~Pires}, \citenamefont {Queiroz},\ and\ \citenamefont {Rodrigues~da
  Silva}}]{Alves:2011kc}%
  \BibitemOpen
  \bibfield  {author} {\bibinfo {author} {\bibfnamefont {A.}~\bibnamefont
  {Alves}}, \bibinfo {author} {\bibfnamefont {E.}~\bibnamefont
  {Ramirez~Barreto}}, \bibinfo {author} {\bibfnamefont {A.~G.}\ \bibnamefont
  {Dias}}, \bibinfo {author} {\bibfnamefont {C.~A.}\ \bibnamefont
  {de~S.~Pires}}, \bibinfo {author} {\bibfnamefont {F.~S.}\ \bibnamefont
  {Queiroz}}, \ and\ \bibinfo {author} {\bibfnamefont {P.~S.}\ \bibnamefont
  {Rodrigues~da Silva}},\ }\href {\doibase 10.1103/PhysRevD.84.115004}
  {\bibfield  {journal} {\bibinfo  {journal} {Phys. Rev.}\ }\textbf {\bibinfo
  {volume} {D84}},\ \bibinfo {pages} {115004} (\bibinfo {year} {2011})},\
  \Eprint {http://arxiv.org/abs/1109.0238} {arXiv:1109.0238 [hep-ph]}
  \BibitemShut {NoStop}%
%%CITATION = ARXIV:1109.0238;%%
\bibitem [{\citenamefont {Alves}\ \emph {et~al.}(2013)\citenamefont {Alves},
  \citenamefont {Ramirez~Barreto}, \citenamefont {Dias}, \citenamefont
  {de~S.~Pires}, \citenamefont {Queiroz},\ and\ \citenamefont {Rodrigues~da
  Silva}}]{Alves:2012yp}%
  \BibitemOpen
  \bibfield  {author} {\bibinfo {author} {\bibfnamefont {A.}~\bibnamefont
  {Alves}}, \bibinfo {author} {\bibfnamefont {E.}~\bibnamefont
  {Ramirez~Barreto}}, \bibinfo {author} {\bibfnamefont {A.~G.}\ \bibnamefont
  {Dias}}, \bibinfo {author} {\bibfnamefont {C.~A.}\ \bibnamefont
  {de~S.~Pires}}, \bibinfo {author} {\bibfnamefont {F.~S.}\ \bibnamefont
  {Queiroz}}, \ and\ \bibinfo {author} {\bibfnamefont {P.~S.}\ \bibnamefont
  {Rodrigues~da Silva}},\ }\href {\doibase 10.1140/epjc/s10052-013-2288-y}
  {\bibfield  {journal} {\bibinfo  {journal} {Eur. Phys. J.}\ }\textbf
  {\bibinfo {volume} {C73}},\ \bibinfo {pages} {2288} (\bibinfo {year}
  {2013})},\ \Eprint {http://arxiv.org/abs/1207.3699} {arXiv:1207.3699
  [hep-ph]} \BibitemShut {NoStop}%
%%CITATION = ARXIV:1207.3699;%%
\bibitem [{\citenamefont {Machado}\ \emph {et~al.}(2018)\citenamefont
  {Machado}, \citenamefont {Pasquini},\ and\ \citenamefont
  {Pleitez}}]{Machado:2018sfh}%
  \BibitemOpen
  \bibfield  {author} {\bibinfo {author} {\bibfnamefont {A.~C.~B.}\
  \bibnamefont {Machado}}, \bibinfo {author} {\bibfnamefont {P.}~\bibnamefont
  {Pasquini}}, \ and\ \bibinfo {author} {\bibfnamefont {V.}~\bibnamefont
  {Pleitez}},\ }\href@noop {} {\  (\bibinfo {year} {2018})},\ \Eprint
  {http://arxiv.org/abs/1810.02817} {arXiv:1810.02817 [hep-ph]} \BibitemShut
  {NoStop}%
%%CITATION = ARXIV:1810.02817;%%
\bibitem [{\citenamefont {Cárcamo~Hernández}\ \emph
  {et~al.}(2019)\citenamefont {Cárcamo~Hernández}, \citenamefont
  {Hidalgo~Velásquez},\ and\ \citenamefont
  {Pérez-Julve}}]{CarcamoHernandez:2019vih}%
  \BibitemOpen
  \bibfield  {author} {\bibinfo {author} {\bibfnamefont {A.~E.}\ \bibnamefont
  {Cárcamo~Hernández}}, \bibinfo {author} {\bibfnamefont {Y.}~\bibnamefont
  {Hidalgo~Velásquez}}, \ and\ \bibinfo {author} {\bibfnamefont {N.~A.}\
  \bibnamefont {Pérez-Julve}},\ }\href@noop {} {\  (\bibinfo {year} {2019})},\
  \Eprint {http://arxiv.org/abs/1905.02323} {arXiv:1905.02323 [hep-ph]}
  \BibitemShut {NoStop}%
%%CITATION = ARXIV:1905.02323;%%
\bibitem [{\citenamefont {Ferreira}\ \emph {et~al.}(2019)\citenamefont
  {Ferreira}, \citenamefont {de~Melo}, \citenamefont {Kovalenko}, \citenamefont
  {Pinheiro},\ and\ \citenamefont {Queiroz}}]{Ferreira:2019qpf}%
  \BibitemOpen
  \bibfield  {author} {\bibinfo {author} {\bibfnamefont {M.~M.}\ \bibnamefont
  {Ferreira}}, \bibinfo {author} {\bibfnamefont {T.~B.}\ \bibnamefont
  {de~Melo}}, \bibinfo {author} {\bibfnamefont {S.}~\bibnamefont {Kovalenko}},
  \bibinfo {author} {\bibfnamefont {P.~R.~D.}\ \bibnamefont {Pinheiro}}, \ and\
  \bibinfo {author} {\bibfnamefont {F.~S.}\ \bibnamefont {Queiroz}},\
  }\href@noop {} {\  (\bibinfo {year} {2019})},\ \Eprint
  {http://arxiv.org/abs/1903.07634} {arXiv:1903.07634 [hep-ph]} \BibitemShut
  {NoStop}%
%%CITATION = ARXIV:1903.07634;%%
\bibitem [{\citenamefont {Gunion}\ \emph {et~al.}(1990)\citenamefont {Gunion},
  \citenamefont {Vega},\ and\ \citenamefont {Wudka}}]{Gunion:1989ci}%
  \BibitemOpen
  \bibfield  {author} {\bibinfo {author} {\bibfnamefont {J.~F.}\ \bibnamefont
  {Gunion}}, \bibinfo {author} {\bibfnamefont {R.}~\bibnamefont {Vega}}, \ and\
  \bibinfo {author} {\bibfnamefont {J.}~\bibnamefont {Wudka}},\ }\href
  {\doibase 10.1103/PhysRevD.42.1673} {\bibfield  {journal} {\bibinfo
  {journal} {Phys. Rev.}\ }\textbf {\bibinfo {volume} {D42}},\ \bibinfo {pages}
  {1673} (\bibinfo {year} {1990})}\BibitemShut {NoStop}%
%%CITATION = PHRVA,D42,1673;%%
\bibitem [{\citenamefont {Akeroyd}\ and\ \citenamefont
  {Moretti}(2012)}]{Akeroyd:2012ms}%
  \BibitemOpen
  \bibfield  {author} {\bibinfo {author} {\bibfnamefont {A.~G.}\ \bibnamefont
  {Akeroyd}}\ and\ \bibinfo {author} {\bibfnamefont {S.}~\bibnamefont
  {Moretti}},\ }\href {\doibase 10.1103/PhysRevD.86.035015} {\bibfield
  {journal} {\bibinfo  {journal} {Phys. Rev.}\ }\textbf {\bibinfo {volume}
  {D86}},\ \bibinfo {pages} {035015} (\bibinfo {year} {2012})},\ \Eprint
  {http://arxiv.org/abs/1206.0535} {arXiv:1206.0535 [hep-ph]} \BibitemShut
  {NoStop}%
%%CITATION = ARXIV:1206.0535;%%
\bibitem [{\citenamefont {de~Medeiros~Varzielas}\ \emph
  {et~al.}(2017)\citenamefont {de~Medeiros~Varzielas}, \citenamefont {King},
  \citenamefont {Luhn},\ and\ \citenamefont
  {Neder}}]{deMedeirosVarzielas:2017glw}%
  \BibitemOpen
  \bibfield  {author} {\bibinfo {author} {\bibfnamefont {I.}~\bibnamefont
  {de~Medeiros~Varzielas}}, \bibinfo {author} {\bibfnamefont {S.~F.}\
  \bibnamefont {King}}, \bibinfo {author} {\bibfnamefont {C.}~\bibnamefont
  {Luhn}}, \ and\ \bibinfo {author} {\bibfnamefont {T.}~\bibnamefont {Neder}},\
  }\href {\doibase 10.1016/j.physletb.2017.11.005} {\bibfield  {journal}
  {\bibinfo  {journal} {Phys. Lett.}\ }\textbf {\bibinfo {volume} {B775}},\
  \bibinfo {pages} {303} (\bibinfo {year} {2017})},\ \Eprint
  {http://arxiv.org/abs/1704.06322} {arXiv:1704.06322 [hep-ph]} \BibitemShut
  {NoStop}%
%%CITATION = ARXIV:1704.06322;%%
\bibitem [{\citenamefont {Kumar~Ghosh}\ \emph {et~al.}(2019)\citenamefont
  {Kumar~Ghosh}, \citenamefont {Ghosh},\ and\ \citenamefont
  {Mukhopadhyaya}}]{Ghosh:2018jpa}%
  \BibitemOpen
  \bibfield  {author} {\bibinfo {author} {\bibfnamefont {D.}~\bibnamefont
  {Kumar~Ghosh}}, \bibinfo {author} {\bibfnamefont {N.}~\bibnamefont {Ghosh}},
  \ and\ \bibinfo {author} {\bibfnamefont {B.}~\bibnamefont {Mukhopadhyaya}},\
  }\href {\doibase 10.1103/PhysRevD.99.015036} {\bibfield  {journal} {\bibinfo
  {journal} {Phys. Rev.}\ }\textbf {\bibinfo {volume} {D99}},\ \bibinfo {pages}
  {015036} (\bibinfo {year} {2019})},\ \Eprint
  {http://arxiv.org/abs/1808.01775} {arXiv:1808.01775 [hep-ph]} \BibitemShut
  {NoStop}%
%%CITATION = ARXIV:1808.01775;%%
\bibitem [{\citenamefont {Chala}\ \emph {et~al.}(2019)\citenamefont {Chala},
  \citenamefont {Ramos},\ and\ \citenamefont {Spannowsky}}]{Chala:2018opy}%
  \BibitemOpen
  \bibfield  {author} {\bibinfo {author} {\bibfnamefont {M.}~\bibnamefont
  {Chala}}, \bibinfo {author} {\bibfnamefont {M.}~\bibnamefont {Ramos}}, \ and\
  \bibinfo {author} {\bibfnamefont {M.}~\bibnamefont {Spannowsky}},\ }\href
  {\doibase 10.1140/epjc/s10052-019-6655-1} {\bibfield  {journal} {\bibinfo
  {journal} {Eur. Phys. J.}\ }\textbf {\bibinfo {volume} {C79}},\ \bibinfo
  {pages} {156} (\bibinfo {year} {2019})},\ \Eprint
  {http://arxiv.org/abs/1812.01901} {arXiv:1812.01901 [hep-ph]} \BibitemShut
  {NoStop}%
%%CITATION = ARXIV:1812.01901;%%
\bibitem [{\citenamefont {Chabab}\ \emph {et~al.}(2018)\citenamefont {Chabab},
  \citenamefont {Peyranère},\ and\ \citenamefont {Rahili}}]{Chabab:2018ert}%
  \BibitemOpen
  \bibfield  {author} {\bibinfo {author} {\bibfnamefont {M.}~\bibnamefont
  {Chabab}}, \bibinfo {author} {\bibfnamefont {M.~C.}\ \bibnamefont
  {Peyranère}}, \ and\ \bibinfo {author} {\bibfnamefont {L.}~\bibnamefont
  {Rahili}},\ }\href {\doibase 10.1140/epjc/s10052-018-6339-2} {\bibfield
  {journal} {\bibinfo  {journal} {Eur. Phys. J.}\ }\textbf {\bibinfo {volume}
  {C78}},\ \bibinfo {pages} {873} (\bibinfo {year} {2018})},\ \Eprint
  {http://arxiv.org/abs/1805.00286} {arXiv:1805.00286 [hep-ph]} \BibitemShut
  {NoStop}%
%%CITATION = ARXIV:1805.00286;%%
\bibitem [{\citenamefont {Chakraborty}\ \emph {et~al.}(2019)\citenamefont
  {Chakraborty}, \citenamefont {Parida},\ and\ \citenamefont
  {Sahoo}}]{Chakraborty:2019uxk}%
  \BibitemOpen
  \bibfield  {author} {\bibinfo {author} {\bibfnamefont {M.}~\bibnamefont
  {Chakraborty}}, \bibinfo {author} {\bibfnamefont {M.~K.}\ \bibnamefont
  {Parida}}, \ and\ \bibinfo {author} {\bibfnamefont {B.}~\bibnamefont
  {Sahoo}},\ }\href@noop {} {\  (\bibinfo {year} {2019})},\ \Eprint
  {http://arxiv.org/abs/1906.05601} {arXiv:1906.05601 [hep-ph]} \BibitemShut
  {NoStop}%
%%CITATION = ARXIV:1906.05601;%%
\bibitem [{\citenamefont {Arkani-Hamed}\ \emph {et~al.}(2002)\citenamefont
  {Arkani-Hamed}, \citenamefont {Cohen}, \citenamefont {Katz}, \citenamefont
  {Nelson}, \citenamefont {Gregoire},\ and\ \citenamefont
  {Wacker}}]{ArkaniHamed:2002qx}%
  \BibitemOpen
  \bibfield  {author} {\bibinfo {author} {\bibfnamefont {N.}~\bibnamefont
  {Arkani-Hamed}}, \bibinfo {author} {\bibfnamefont {A.~G.}\ \bibnamefont
  {Cohen}}, \bibinfo {author} {\bibfnamefont {E.}~\bibnamefont {Katz}},
  \bibinfo {author} {\bibfnamefont {A.~E.}\ \bibnamefont {Nelson}}, \bibinfo
  {author} {\bibfnamefont {T.}~\bibnamefont {Gregoire}}, \ and\ \bibinfo
  {author} {\bibfnamefont {J.~G.}\ \bibnamefont {Wacker}},\ }\href {\doibase
  10.1088/1126-6708/2002/08/021} {\bibfield  {journal} {\bibinfo  {journal}
  {JHEP}\ }\textbf {\bibinfo {volume} {08}},\ \bibinfo {pages} {021} (\bibinfo
  {year} {2002})},\ \Eprint {http://arxiv.org/abs/hep-ph/0206020}
  {arXiv:hep-ph/0206020 [hep-ph]} \BibitemShut {NoStop}%
%%CITATION = HEP-PH/0206020;%%
\bibitem [{\citenamefont {Georgi}\ and\ \citenamefont
  {Machacek}(1985)}]{Georgi:1985nv}%
  \BibitemOpen
  \bibfield  {author} {\bibinfo {author} {\bibfnamefont {H.}~\bibnamefont
  {Georgi}}\ and\ \bibinfo {author} {\bibfnamefont {M.}~\bibnamefont
  {Machacek}},\ }\href {\doibase 10.1016/0550-3213(85)90325-6} {\bibfield
  {journal} {\bibinfo  {journal} {Nucl. Phys.}\ }\textbf {\bibinfo {volume}
  {B262}},\ \bibinfo {pages} {463} (\bibinfo {year} {1985})}\BibitemShut
  {NoStop}%
%%CITATION = NUPHA,B262,463;%%
\bibitem [{\citenamefont {Sun}\ \emph {et~al.}(2017)\citenamefont {Sun},
  \citenamefont {Luo}, \citenamefont {Wei},\ and\ \citenamefont
  {Liu}}]{Sun:2017mue}%
  \BibitemOpen
  \bibfield  {author} {\bibinfo {author} {\bibfnamefont {H.}~\bibnamefont
  {Sun}}, \bibinfo {author} {\bibfnamefont {X.}~\bibnamefont {Luo}}, \bibinfo
  {author} {\bibfnamefont {W.}~\bibnamefont {Wei}}, \ and\ \bibinfo {author}
  {\bibfnamefont {T.}~\bibnamefont {Liu}},\ }\href {\doibase
  10.1103/PhysRevD.96.095003} {\bibfield  {journal} {\bibinfo  {journal} {Phys.
  Rev.}\ }\textbf {\bibinfo {volume} {D96}},\ \bibinfo {pages} {095003}
  (\bibinfo {year} {2017})},\ \Eprint {http://arxiv.org/abs/1710.06284}
  {arXiv:1710.06284 [hep-ph]} \BibitemShut {NoStop}%
%%CITATION = ARXIV:1710.06284;%%
\bibitem [{\citenamefont {Chiang}\ \emph {et~al.}(2019)\citenamefont {Chiang},
  \citenamefont {Cottin},\ and\ \citenamefont {Eberhardt}}]{Chiang:2018cgb}%
  \BibitemOpen
  \bibfield  {author} {\bibinfo {author} {\bibfnamefont {C.-W.}\ \bibnamefont
  {Chiang}}, \bibinfo {author} {\bibfnamefont {G.}~\bibnamefont {Cottin}}, \
  and\ \bibinfo {author} {\bibfnamefont {O.}~\bibnamefont {Eberhardt}},\ }\href
  {\doibase 10.1103/PhysRevD.99.015001} {\bibfield  {journal} {\bibinfo
  {journal} {Phys. Rev.}\ }\textbf {\bibinfo {volume} {D99}},\ \bibinfo {pages}
  {015001} (\bibinfo {year} {2019})},\ \Eprint
  {http://arxiv.org/abs/1807.10660} {arXiv:1807.10660 [hep-ph]} \BibitemShut
  {NoStop}%
%%CITATION = ARXIV:1807.10660;%%
\bibitem [{\citenamefont {Zee}(1986)}]{Zee:1985id}%
  \BibitemOpen
  \bibfield  {author} {\bibinfo {author} {\bibfnamefont {A.}~\bibnamefont
  {Zee}},\ }\href {\doibase 10.1016/0550-3213(86)90475-X} {\bibfield  {journal}
  {\bibinfo  {journal} {Nucl. Phys.}\ }\textbf {\bibinfo {volume} {B264}},\
  \bibinfo {pages} {99} (\bibinfo {year} {1986})}\BibitemShut {NoStop}%
%%CITATION = NUPHA,B264,99;%%
\bibitem [{\citenamefont {Nebot}\ \emph {et~al.}(2008)\citenamefont {Nebot},
  \citenamefont {Oliver}, \citenamefont {Palao},\ and\ \citenamefont
  {Santamaria}}]{Nebot:2007bc}%
  \BibitemOpen
  \bibfield  {author} {\bibinfo {author} {\bibfnamefont {M.}~\bibnamefont
  {Nebot}}, \bibinfo {author} {\bibfnamefont {J.~F.}\ \bibnamefont {Oliver}},
  \bibinfo {author} {\bibfnamefont {D.}~\bibnamefont {Palao}}, \ and\ \bibinfo
  {author} {\bibfnamefont {A.}~\bibnamefont {Santamaria}},\ }\href {\doibase
  10.1103/PhysRevD.77.093013} {\bibfield  {journal} {\bibinfo  {journal} {Phys.
  Rev.}\ }\textbf {\bibinfo {volume} {D77}},\ \bibinfo {pages} {093013}
  (\bibinfo {year} {2008})},\ \Eprint {http://arxiv.org/abs/0711.0483}
  {arXiv:0711.0483 [hep-ph]} \BibitemShut {NoStop}%
%%CITATION = ARXIV:0711.0483;%%
\bibitem [{\citenamefont {Camargo}\ \emph {et~al.}(2019)\citenamefont
  {Camargo}, \citenamefont {Dias}, \citenamefont {de~Melo},\ and\ \citenamefont
  {Queiroz}}]{Camargo:2018uzw}%
  \BibitemOpen
  \bibfield  {author} {\bibinfo {author} {\bibfnamefont {D.~A.}\ \bibnamefont
  {Camargo}}, \bibinfo {author} {\bibfnamefont {A.~G.}\ \bibnamefont {Dias}},
  \bibinfo {author} {\bibfnamefont {T.~B.}\ \bibnamefont {de~Melo}}, \ and\
  \bibinfo {author} {\bibfnamefont {F.~S.}\ \bibnamefont {Queiroz}},\ }\href
  {\doibase 10.1007/JHEP04(2019)129} {\bibfield  {journal} {\bibinfo  {journal}
  {JHEP}\ }\textbf {\bibinfo {volume} {04}},\ \bibinfo {pages} {129} (\bibinfo
  {year} {2019})},\ \Eprint {http://arxiv.org/abs/1811.05488} {arXiv:1811.05488
  [hep-ph]} \BibitemShut {NoStop}%
%%CITATION = ARXIV:1811.05488;%%
\bibitem [{\citenamefont {Mitra}\ \emph {et~al.}(2017)\citenamefont {Mitra},
  \citenamefont {Niyogi},\ and\ \citenamefont {Spannowsky}}]{Mitra:2016wpr}%
  \BibitemOpen
  \bibfield  {author} {\bibinfo {author} {\bibfnamefont {M.}~\bibnamefont
  {Mitra}}, \bibinfo {author} {\bibfnamefont {S.}~\bibnamefont {Niyogi}}, \
  and\ \bibinfo {author} {\bibfnamefont {M.}~\bibnamefont {Spannowsky}},\
  }\href {\doibase 10.1103/PhysRevD.95.035042} {\bibfield  {journal} {\bibinfo
  {journal} {Phys. Rev.}\ }\textbf {\bibinfo {volume} {D95}},\ \bibinfo {pages}
  {035042} (\bibinfo {year} {2017})},\ \Eprint
  {http://arxiv.org/abs/1611.09594} {arXiv:1611.09594 [hep-ph]} \BibitemShut
  {NoStop}%
%%CITATION = ARXIV:1611.09594;%%
\bibitem [{\citenamefont {Du}\ \emph {et~al.}(2019)\citenamefont {Du},
  \citenamefont {Dunbrack}, \citenamefont {Ramsey-Musolf},\ and\ \citenamefont
  {Yu}}]{Du:2018eaw}%
  \BibitemOpen
  \bibfield  {author} {\bibinfo {author} {\bibfnamefont {Y.}~\bibnamefont
  {Du}}, \bibinfo {author} {\bibfnamefont {A.}~\bibnamefont {Dunbrack}},
  \bibinfo {author} {\bibfnamefont {M.~J.}\ \bibnamefont {Ramsey-Musolf}}, \
  and\ \bibinfo {author} {\bibfnamefont {J.-H.}\ \bibnamefont {Yu}},\ }\href
  {\doibase 10.1007/JHEP01(2019)101} {\bibfield  {journal} {\bibinfo  {journal}
  {JHEP}\ }\textbf {\bibinfo {volume} {01}},\ \bibinfo {pages} {101} (\bibinfo
  {year} {2019})},\ \Eprint {http://arxiv.org/abs/1810.09450} {arXiv:1810.09450
  [hep-ph]} \BibitemShut {NoStop}%
%%CITATION = ARXIV:1810.09450;%%
\bibitem [{\citenamefont {Huitu}\ \emph {et~al.}(1997)\citenamefont {Huitu},
  \citenamefont {Maalampi}, \citenamefont {Pietila},\ and\ \citenamefont
  {Raidal}}]{Huitu:1996su}%
  \BibitemOpen
  \bibfield  {author} {\bibinfo {author} {\bibfnamefont {K.}~\bibnamefont
  {Huitu}}, \bibinfo {author} {\bibfnamefont {J.}~\bibnamefont {Maalampi}},
  \bibinfo {author} {\bibfnamefont {A.}~\bibnamefont {Pietila}}, \ and\
  \bibinfo {author} {\bibfnamefont {M.}~\bibnamefont {Raidal}},\ }\href
  {\doibase 10.1016/S0550-3213(97)87466-4} {\bibfield  {journal} {\bibinfo
  {journal} {Nucl. Phys.}\ }\textbf {\bibinfo {volume} {B487}},\ \bibinfo
  {pages} {27} (\bibinfo {year} {1997})},\ \Eprint
  {http://arxiv.org/abs/hep-ph/9606311} {arXiv:hep-ph/9606311 [hep-ph]}
  \BibitemShut {NoStop}%
%%CITATION = HEP-PH/9606311;%%
\bibitem [{\citenamefont {Han}\ \emph {et~al.}(2007)\citenamefont {Han},
  \citenamefont {Mukhopadhyaya}, \citenamefont {Si},\ and\ \citenamefont
  {Wang}}]{Han:2007bk}%
  \BibitemOpen
  \bibfield  {author} {\bibinfo {author} {\bibfnamefont {T.}~\bibnamefont
  {Han}}, \bibinfo {author} {\bibfnamefont {B.}~\bibnamefont {Mukhopadhyaya}},
  \bibinfo {author} {\bibfnamefont {Z.}~\bibnamefont {Si}}, \ and\ \bibinfo
  {author} {\bibfnamefont {K.}~\bibnamefont {Wang}},\ }\href {\doibase
  10.1103/PhysRevD.76.075013} {\bibfield  {journal} {\bibinfo  {journal} {Phys.
  Rev.}\ }\textbf {\bibinfo {volume} {D76}},\ \bibinfo {pages} {075013}
  (\bibinfo {year} {2007})},\ \Eprint {http://arxiv.org/abs/0706.0441}
  {arXiv:0706.0441 [hep-ph]} \BibitemShut {NoStop}%
%%CITATION = ARXIV:0706.0441;%%
\bibitem [{\citenamefont {Chao}\ \emph {et~al.}(2008)\citenamefont {Chao},
  \citenamefont {Si}, \citenamefont {Xing},\ and\ \citenamefont
  {Zhou}}]{Chao:2008mq}%
  \BibitemOpen
  \bibfield  {author} {\bibinfo {author} {\bibfnamefont {W.}~\bibnamefont
  {Chao}}, \bibinfo {author} {\bibfnamefont {Z.-G.}\ \bibnamefont {Si}},
  \bibinfo {author} {\bibfnamefont {Z.-z.}\ \bibnamefont {Xing}}, \ and\
  \bibinfo {author} {\bibfnamefont {S.}~\bibnamefont {Zhou}},\ }\href {\doibase
  10.1016/j.physletb.2008.08.003} {\bibfield  {journal} {\bibinfo  {journal}
  {Phys. Lett.}\ }\textbf {\bibinfo {volume} {B666}},\ \bibinfo {pages} {451}
  (\bibinfo {year} {2008})},\ \Eprint {http://arxiv.org/abs/0804.1265}
  {arXiv:0804.1265 [hep-ph]} \BibitemShut {NoStop}%
%%CITATION = ARXIV:0804.1265;%%
\bibitem [{\citenamefont {Akeroyd}\ \emph {et~al.}(2010)\citenamefont
  {Akeroyd}, \citenamefont {Chiang},\ and\ \citenamefont
  {Gaur}}]{Akeroyd:2010ip}%
  \BibitemOpen
  \bibfield  {author} {\bibinfo {author} {\bibfnamefont {A.~G.}\ \bibnamefont
  {Akeroyd}}, \bibinfo {author} {\bibfnamefont {C.-W.}\ \bibnamefont {Chiang}},
  \ and\ \bibinfo {author} {\bibfnamefont {N.}~\bibnamefont {Gaur}},\ }\href
  {\doibase 10.1007/JHEP11(2010)005} {\bibfield  {journal} {\bibinfo  {journal}
  {JHEP}\ }\textbf {\bibinfo {volume} {11}},\ \bibinfo {pages} {005} (\bibinfo
  {year} {2010})},\ \Eprint {http://arxiv.org/abs/1009.2780} {arXiv:1009.2780
  [hep-ph]} \BibitemShut {NoStop}%
%%CITATION = ARXIV:1009.2780;%%
\bibitem [{\citenamefont {Chiang}\ \emph {et~al.}(2012)\citenamefont {Chiang},
  \citenamefont {Nomura},\ and\ \citenamefont {Tsumura}}]{Chiang:2012dk}%
  \BibitemOpen
  \bibfield  {author} {\bibinfo {author} {\bibfnamefont {C.-W.}\ \bibnamefont
  {Chiang}}, \bibinfo {author} {\bibfnamefont {T.}~\bibnamefont {Nomura}}, \
  and\ \bibinfo {author} {\bibfnamefont {K.}~\bibnamefont {Tsumura}},\ }\href
  {\doibase 10.1103/PhysRevD.85.095023} {\bibfield  {journal} {\bibinfo
  {journal} {Phys. Rev.}\ }\textbf {\bibinfo {volume} {D85}},\ \bibinfo {pages}
  {095023} (\bibinfo {year} {2012})},\ \Eprint {http://arxiv.org/abs/1202.2014}
  {arXiv:1202.2014 [hep-ph]} \BibitemShut {NoStop}%
%%CITATION = ARXIV:1202.2014;%%
\bibitem [{\citenamefont {del Águila}\ and\ \citenamefont
  {Chala}(2014)}]{delAguila:2013mia}%
  \BibitemOpen
  \bibfield  {author} {\bibinfo {author} {\bibfnamefont {F.}~\bibnamefont {del
  Águila}}\ and\ \bibinfo {author} {\bibfnamefont {M.}~\bibnamefont {Chala}},\
  }\href {\doibase 10.1007/JHEP03(2014)027} {\bibfield  {journal} {\bibinfo
  {journal} {JHEP}\ }\textbf {\bibinfo {volume} {03}},\ \bibinfo {pages} {027}
  (\bibinfo {year} {2014})},\ \Eprint {http://arxiv.org/abs/1311.1510}
  {arXiv:1311.1510 [hep-ph]} \BibitemShut {NoStop}%
%%CITATION = ARXIV:1311.1510;%%
\bibitem [{\citenamefont {Kanemura}\ \emph {et~al.}(2013)\citenamefont
  {Kanemura}, \citenamefont {Yagyu},\ and\ \citenamefont
  {Yokoya}}]{Kanemura:2013vxa}%
  \BibitemOpen
  \bibfield  {author} {\bibinfo {author} {\bibfnamefont {S.}~\bibnamefont
  {Kanemura}}, \bibinfo {author} {\bibfnamefont {K.}~\bibnamefont {Yagyu}}, \
  and\ \bibinfo {author} {\bibfnamefont {H.}~\bibnamefont {Yokoya}},\ }\href
  {\doibase 10.1016/j.physletb.2013.08.054} {\bibfield  {journal} {\bibinfo
  {journal} {Phys. Lett.}\ }\textbf {\bibinfo {volume} {B726}},\ \bibinfo
  {pages} {316} (\bibinfo {year} {2013})},\ \Eprint
  {http://arxiv.org/abs/1305.2383} {arXiv:1305.2383 [hep-ph]} \BibitemShut
  {NoStop}%
%%CITATION = ARXIV:1305.2383;%%
\bibitem [{\citenamefont {King}\ \emph {et~al.}(2014)\citenamefont {King},
  \citenamefont {Merle},\ and\ \citenamefont {Panizzi}}]{King:2014uha}%
  \BibitemOpen
  \bibfield  {author} {\bibinfo {author} {\bibfnamefont {S.~F.}\ \bibnamefont
  {King}}, \bibinfo {author} {\bibfnamefont {A.}~\bibnamefont {Merle}}, \ and\
  \bibinfo {author} {\bibfnamefont {L.}~\bibnamefont {Panizzi}},\ }\href
  {\doibase 10.1007/JHEP11(2014)124} {\bibfield  {journal} {\bibinfo  {journal}
  {JHEP}\ }\textbf {\bibinfo {volume} {11}},\ \bibinfo {pages} {124} (\bibinfo
  {year} {2014})},\ \Eprint {http://arxiv.org/abs/1406.4137} {arXiv:1406.4137
  [hep-ph]} \BibitemShut {NoStop}%
%%CITATION = ARXIV:1406.4137;%%
\bibitem [{\citenamefont {Kang}\ \emph {et~al.}(2015)\citenamefont {Kang},
  \citenamefont {Li}, \citenamefont {Li}, \citenamefont {Liu},\ and\
  \citenamefont {Ning}}]{kang:2014jia}%
  \BibitemOpen
  \bibfield  {author} {\bibinfo {author} {\bibfnamefont {Z.}~\bibnamefont
  {Kang}}, \bibinfo {author} {\bibfnamefont {J.}~\bibnamefont {Li}}, \bibinfo
  {author} {\bibfnamefont {T.}~\bibnamefont {Li}}, \bibinfo {author}
  {\bibfnamefont {Y.}~\bibnamefont {Liu}}, \ and\ \bibinfo {author}
  {\bibfnamefont {G.-Z.}\ \bibnamefont {Ning}},\ }\href {\doibase
  10.1140/epjc/s10052-015-3774-1} {\bibfield  {journal} {\bibinfo  {journal}
  {Eur. Phys. J.}\ }\textbf {\bibinfo {volume} {C75}},\ \bibinfo {pages} {574}
  (\bibinfo {year} {2015})},\ \Eprint {http://arxiv.org/abs/1404.5207}
  {arXiv:1404.5207 [hep-ph]} \BibitemShut {NoStop}%
%%CITATION = ARXIV:1404.5207;%%
\bibitem [{\citenamefont {Kanemura}\ \emph {et~al.}(2015)\citenamefont
  {Kanemura}, \citenamefont {Kikuchi}, \citenamefont {Yokoya},\ and\
  \citenamefont {Yagyu}}]{Kanemura:2014ipa}%
  \BibitemOpen
  \bibfield  {author} {\bibinfo {author} {\bibfnamefont {S.}~\bibnamefont
  {Kanemura}}, \bibinfo {author} {\bibfnamefont {M.}~\bibnamefont {Kikuchi}},
  \bibinfo {author} {\bibfnamefont {H.}~\bibnamefont {Yokoya}}, \ and\ \bibinfo
  {author} {\bibfnamefont {K.}~\bibnamefont {Yagyu}},\ }\href {\doibase
  10.1093/ptep/ptv071} {\bibfield  {journal} {\bibinfo  {journal} {PTEP}\
  }\textbf {\bibinfo {volume} {2015}},\ \bibinfo {pages} {051B02} (\bibinfo
  {year} {2015})},\ \Eprint {http://arxiv.org/abs/1412.7603} {arXiv:1412.7603
  [hep-ph]} \BibitemShut {NoStop}%
%%CITATION = ARXIV:1412.7603;%%
\bibitem [{\citenamefont {Bambhaniya}\ \emph {et~al.}(2015)\citenamefont
  {Bambhaniya}, \citenamefont {Chakrabortty}, \citenamefont {Gluza},
  \citenamefont {Jelinski},\ and\ \citenamefont
  {Szafron}}]{Bambhaniya:2015wna}%
  \BibitemOpen
  \bibfield  {author} {\bibinfo {author} {\bibfnamefont {G.}~\bibnamefont
  {Bambhaniya}}, \bibinfo {author} {\bibfnamefont {J.}~\bibnamefont
  {Chakrabortty}}, \bibinfo {author} {\bibfnamefont {J.}~\bibnamefont {Gluza}},
  \bibinfo {author} {\bibfnamefont {T.}~\bibnamefont {Jelinski}}, \ and\
  \bibinfo {author} {\bibfnamefont {R.}~\bibnamefont {Szafron}},\ }\href
  {\doibase 10.1103/PhysRevD.92.015016} {\bibfield  {journal} {\bibinfo
  {journal} {Phys. Rev.}\ }\textbf {\bibinfo {volume} {D92}},\ \bibinfo {pages}
  {015016} (\bibinfo {year} {2015})},\ \Eprint
  {http://arxiv.org/abs/1504.03999} {arXiv:1504.03999 [hep-ph]} \BibitemShut
  {NoStop}%
%%CITATION = ARXIV:1504.03999;%%
\bibitem [{\citenamefont {Fileviez~Perez}\ \emph
  {et~al.}(2008{\natexlab{b}})\citenamefont {Fileviez~Perez}, \citenamefont
  {Han}, \citenamefont {Huang}, \citenamefont {Li},\ and\ \citenamefont
  {Wang}}]{Perez:2008ha}%
  \BibitemOpen
  \bibfield  {author} {\bibinfo {author} {\bibfnamefont {P.}~\bibnamefont
  {Fileviez~Perez}}, \bibinfo {author} {\bibfnamefont {T.}~\bibnamefont {Han}},
  \bibinfo {author} {\bibfnamefont {G.-y.}\ \bibnamefont {Huang}}, \bibinfo
  {author} {\bibfnamefont {T.}~\bibnamefont {Li}}, \ and\ \bibinfo {author}
  {\bibfnamefont {K.}~\bibnamefont {Wang}},\ }\href {\doibase
  10.1103/PhysRevD.78.015018} {\bibfield  {journal} {\bibinfo  {journal} {Phys.
  Rev.}\ }\textbf {\bibinfo {volume} {D78}},\ \bibinfo {pages} {015018}
  (\bibinfo {year} {2008}{\natexlab{b}})},\ \Eprint
  {http://arxiv.org/abs/0805.3536} {arXiv:0805.3536 [hep-ph]} \BibitemShut
  {NoStop}%
%%CITATION = ARXIV:0805.3536;%%
\bibitem [{\citenamefont {Collaboration}(2012)}]{CMS:2012kua}%
  \BibitemOpen
  \bibfield  {author} {\bibinfo {author} {\bibfnamefont {C.}~\bibnamefont
  {Collaboration}} (\bibinfo {collaboration} {CMS}),\ }\href@noop {} {\
  (\bibinfo {year} {2012})}\BibitemShut {NoStop}%
%%CITATION = CMS-PAS-HIG-12-005;%%
\bibitem [{\citenamefont {Aad}\ \emph {et~al.}(2012)\citenamefont {Aad} \emph
  {et~al.}}]{ATLAS:2012hi}%
  \BibitemOpen
  \bibfield  {author} {\bibinfo {author} {\bibfnamefont {G.}~\bibnamefont
  {Aad}} \emph {et~al.} (\bibinfo {collaboration} {ATLAS}),\ }\href {\doibase
  10.1140/epjc/s10052-012-2244-2} {\bibfield  {journal} {\bibinfo  {journal}
  {Eur. Phys. J.}\ }\textbf {\bibinfo {volume} {C72}},\ \bibinfo {pages} {2244}
  (\bibinfo {year} {2012})},\ \Eprint {http://arxiv.org/abs/1210.5070}
  {arXiv:1210.5070 [hep-ex]} \BibitemShut {NoStop}%
%%CITATION = ARXIV:1210.5070;%%
\bibitem [{\citenamefont {Chatrchyan}\ \emph {et~al.}(2012)\citenamefont
  {Chatrchyan} \emph {et~al.}}]{Chatrchyan:2012ya}%
  \BibitemOpen
  \bibfield  {author} {\bibinfo {author} {\bibfnamefont {S.}~\bibnamefont
  {Chatrchyan}} \emph {et~al.} (\bibinfo {collaboration} {CMS}),\ }\href
  {\doibase 10.1140/epjc/s10052-012-2189-5} {\bibfield  {journal} {\bibinfo
  {journal} {Eur. Phys. J.}\ }\textbf {\bibinfo {volume} {C72}},\ \bibinfo
  {pages} {2189} (\bibinfo {year} {2012})},\ \Eprint
  {http://arxiv.org/abs/1207.2666} {arXiv:1207.2666 [hep-ex]} \BibitemShut
  {NoStop}%
%%CITATION = ARXIV:1207.2666;%%
\bibitem [{\citenamefont {Aad}\ \emph {et~al.}(2015)\citenamefont {Aad} \emph
  {et~al.}}]{ATLAS:2014kca}%
  \BibitemOpen
  \bibfield  {author} {\bibinfo {author} {\bibfnamefont {G.}~\bibnamefont
  {Aad}} \emph {et~al.} (\bibinfo {collaboration} {ATLAS}),\ }\href {\doibase
  10.1007/JHEP03(2015)041} {\bibfield  {journal} {\bibinfo  {journal} {JHEP}\
  }\textbf {\bibinfo {volume} {03}},\ \bibinfo {pages} {041} (\bibinfo {year}
  {2015})},\ \Eprint {http://arxiv.org/abs/1412.0237} {arXiv:1412.0237
  [hep-ex]} \BibitemShut {NoStop}%
%%CITATION = ARXIV:1412.0237;%%
\bibitem [{\citenamefont {collaboration}(2016)}]{ATLAS:2016pbt}%
  \BibitemOpen
  \bibfield  {author} {\bibinfo {author} {\bibfnamefont {T.~A.}\ \bibnamefont
  {collaboration}} (\bibinfo {collaboration} {ATLAS}),\ }\href@noop {}
  {\bibfield  {journal} {\bibinfo  {journal} {ATLAS-CONF-2016-051}\ } (\bibinfo
  {year} {2016})}\BibitemShut {NoStop}%
%%CITATION = ATLAS-CONF-2016-051;%%
\bibitem [{\citenamefont {Collaboration}(2016)}]{CMS:2016cpz}%
  \BibitemOpen
  \bibfield  {author} {\bibinfo {author} {\bibfnamefont {C.}~\bibnamefont
  {Collaboration}} (\bibinfo {collaboration} {CMS}),\ }\href@noop {} {\bibfield
   {journal} {\bibinfo  {journal} {CMS-PAS-HIG-14-039}\ } (\bibinfo {year}
  {2016})}\BibitemShut {NoStop}%
%%CITATION = CMS-PAS-HIG-14-039;%%
\bibitem [{\citenamefont {Collaboration}(2017)}]{CMS:2017pet}%
  \BibitemOpen
  \bibfield  {author} {\bibinfo {author} {\bibfnamefont {C.}~\bibnamefont
  {Collaboration}} (\bibinfo {collaboration} {CMS}),\ }\href@noop {} {\
  (\bibinfo {year} {2017})}\BibitemShut {NoStop}%
%%CITATION = CMS-PAS-HIG-16-036;%%
\bibitem [{\citenamefont {Aaboud}\ \emph {et~al.}(2018)\citenamefont {Aaboud}
  \emph {et~al.}}]{Aaboud:2017qph}%
  \BibitemOpen
  \bibfield  {author} {\bibinfo {author} {\bibfnamefont {M.}~\bibnamefont
  {Aaboud}} \emph {et~al.} (\bibinfo {collaboration} {ATLAS}),\ }\href
  {\doibase 10.1140/EPJC/S10052-018-5661-Z, 10.1140/epjc/s10052-018-5661-z}
  {\bibfield  {journal} {\bibinfo  {journal} {Eur. Phys. J.}\ }\textbf
  {\bibinfo {volume} {C78}},\ \bibinfo {pages} {199} (\bibinfo {year}
  {2018})},\ \Eprint {http://arxiv.org/abs/1710.09748} {arXiv:1710.09748
  [hep-ex]} \BibitemShut {NoStop}%
%%CITATION = ARXIV:1710.09748;%%
\bibitem [{\citenamefont {Babu}\ and\ \citenamefont
  {Jana}(2017)}]{Babu:2016rcr}%
  \BibitemOpen
  \bibfield  {author} {\bibinfo {author} {\bibfnamefont {K.~S.}\ \bibnamefont
  {Babu}}\ and\ \bibinfo {author} {\bibfnamefont {S.}~\bibnamefont {Jana}},\
  }\href {\doibase 10.1103/PhysRevD.95.055020} {\bibfield  {journal} {\bibinfo
  {journal} {Phys. Rev.}\ }\textbf {\bibinfo {volume} {D95}},\ \bibinfo {pages}
  {055020} (\bibinfo {year} {2017})},\ \Eprint
  {http://arxiv.org/abs/1612.09224} {arXiv:1612.09224 [hep-ph]} \BibitemShut
  {NoStop}%
%%CITATION = ARXIV:1612.09224;%%
\bibitem [{\citenamefont {Crivellin}\ \emph {et~al.}(2019)\citenamefont
  {Crivellin}, \citenamefont {Ghezzi}, \citenamefont {Panizzi}, \citenamefont
  {Pruna},\ and\ \citenamefont {Signer}}]{Crivellin:2018ahj}%
  \BibitemOpen
  \bibfield  {author} {\bibinfo {author} {\bibfnamefont {A.}~\bibnamefont
  {Crivellin}}, \bibinfo {author} {\bibfnamefont {M.}~\bibnamefont {Ghezzi}},
  \bibinfo {author} {\bibfnamefont {L.}~\bibnamefont {Panizzi}}, \bibinfo
  {author} {\bibfnamefont {G.~M.}\ \bibnamefont {Pruna}}, \ and\ \bibinfo
  {author} {\bibfnamefont {A.}~\bibnamefont {Signer}},\ }\href {\doibase
  10.1103/PhysRevD.99.035004} {\bibfield  {journal} {\bibinfo  {journal} {Phys.
  Rev.}\ }\textbf {\bibinfo {volume} {D99}},\ \bibinfo {pages} {035004}
  (\bibinfo {year} {2019})},\ \Eprint {http://arxiv.org/abs/1807.10224}
  {arXiv:1807.10224 [hep-ph]} \BibitemShut {NoStop}%
%%CITATION = ARXIV:1807.10224;%%
\bibitem [{\citenamefont {Li}(2018)}]{Li:2018jns}%
  \BibitemOpen
  \bibfield  {author} {\bibinfo {author} {\bibfnamefont {T.}~\bibnamefont
  {Li}},\ }\href {\doibase 10.1007/JHEP09(2018)079} {\bibfield  {journal}
  {\bibinfo  {journal} {JHEP}\ }\textbf {\bibinfo {volume} {09}},\ \bibinfo
  {pages} {079} (\bibinfo {year} {2018})},\ \Eprint
  {http://arxiv.org/abs/1802.00945} {arXiv:1802.00945 [hep-ph]} \BibitemShut
  {NoStop}%
%%CITATION = ARXIV:1802.00945;%%
\bibitem [{\citenamefont {Mondal}\ and\ \citenamefont
  {Rai}(2016)}]{Mondal:2016czu}%
  \BibitemOpen
  \bibfield  {author} {\bibinfo {author} {\bibfnamefont {S.}~\bibnamefont
  {Mondal}}\ and\ \bibinfo {author} {\bibfnamefont {S.~K.}\ \bibnamefont
  {Rai}},\ }\href {\doibase 10.1103/PhysRevD.93.118702} {\bibfield  {journal}
  {\bibinfo  {journal} {Phys. Rev.}\ }\textbf {\bibinfo {volume} {D93}},\
  \bibinfo {pages} {118702} (\bibinfo {year} {2016})}\BibitemShut {NoStop}%
%%CITATION = PHRVA,D93,118702;%%
\bibitem [{\citenamefont {Queiroz}(2016)}]{Queiroz:2016qmc}%
  \BibitemOpen
  \bibfield  {author} {\bibinfo {author} {\bibfnamefont {F.~S.}\ \bibnamefont
  {Queiroz}},\ }\href {\doibase 10.1103/PhysRevD.93.118701} {\bibfield
  {journal} {\bibinfo  {journal} {Phys. Rev.}\ }\textbf {\bibinfo {volume}
  {D93}},\ \bibinfo {pages} {118701} (\bibinfo {year} {2016})}\BibitemShut
  {NoStop}%
%%CITATION = PHRVA,D93,118701;%%
\bibitem [{\citenamefont {Aad}\ \emph {et~al.}(2016)\citenamefont {Aad} \emph
  {et~al.}}]{Aad:2016jkr}%
  \BibitemOpen
  \bibfield  {author} {\bibinfo {author} {\bibfnamefont {G.}~\bibnamefont
  {Aad}} \emph {et~al.} (\bibinfo {collaboration} {ATLAS}),\ }\href {\doibase
  10.1140/epjc/s10052-016-4120-y} {\bibfield  {journal} {\bibinfo  {journal}
  {Eur. Phys. J.}\ }\textbf {\bibinfo {volume} {C76}},\ \bibinfo {pages} {292}
  (\bibinfo {year} {2016})},\ \Eprint {http://arxiv.org/abs/1603.05598}
  {arXiv:1603.05598 [hep-ex]} \BibitemShut {NoStop}%
%%CITATION = ARXIV:1603.05598;%%
\bibitem [{\citenamefont {Patra}\ \emph {et~al.}(2016)\citenamefont {Patra},
  \citenamefont {Queiroz},\ and\ \citenamefont {Rodejohann}}]{Patra:2015bga}%
  \BibitemOpen
  \bibfield  {author} {\bibinfo {author} {\bibfnamefont {S.}~\bibnamefont
  {Patra}}, \bibinfo {author} {\bibfnamefont {F.~S.}\ \bibnamefont {Queiroz}},
  \ and\ \bibinfo {author} {\bibfnamefont {W.}~\bibnamefont {Rodejohann}},\
  }\href {\doibase 10.1016/j.physletb.2015.11.009} {\bibfield  {journal}
  {\bibinfo  {journal} {Phys. Lett.}\ }\textbf {\bibinfo {volume} {B752}},\
  \bibinfo {pages} {186} (\bibinfo {year} {2016})},\ \Eprint
  {http://arxiv.org/abs/1506.03456} {arXiv:1506.03456 [hep-ph]} \BibitemShut
  {NoStop}%
%%CITATION = ARXIV:1506.03456;%%
\bibitem [{\citenamefont {Lindner}\ \emph
  {et~al.}(2016{\natexlab{a}})\citenamefont {Lindner}, \citenamefont
  {Queiroz},\ and\ \citenamefont {Rodejohann}}]{Lindner:2016lpp}%
  \BibitemOpen
  \bibfield  {author} {\bibinfo {author} {\bibfnamefont {M.}~\bibnamefont
  {Lindner}}, \bibinfo {author} {\bibfnamefont {F.~S.}\ \bibnamefont
  {Queiroz}}, \ and\ \bibinfo {author} {\bibfnamefont {W.}~\bibnamefont
  {Rodejohann}},\ }\href {\doibase 10.1016/j.physletb.2016.08.068} {\bibfield
  {journal} {\bibinfo  {journal} {Phys. Lett.}\ }\textbf {\bibinfo {volume}
  {B762}},\ \bibinfo {pages} {190} (\bibinfo {year} {2016}{\natexlab{a}})},\
  \Eprint {http://arxiv.org/abs/1604.07419} {arXiv:1604.07419 [hep-ph]}
  \BibitemShut {NoStop}%
%%CITATION = ARXIV:1604.07419;%%
\bibitem [{\citenamefont {Antusch}\ \emph {et~al.}(2019)\citenamefont
  {Antusch}, \citenamefont {Fischer}, \citenamefont {Hammad},\ and\
  \citenamefont {Scherb}}]{Antusch:2018svb}%
  \BibitemOpen
  \bibfield  {author} {\bibinfo {author} {\bibfnamefont {S.}~\bibnamefont
  {Antusch}}, \bibinfo {author} {\bibfnamefont {O.}~\bibnamefont {Fischer}},
  \bibinfo {author} {\bibfnamefont {A.}~\bibnamefont {Hammad}}, \ and\ \bibinfo
  {author} {\bibfnamefont {C.}~\bibnamefont {Scherb}},\ }\href {\doibase
  10.1007/JHEP02(2019)157} {\bibfield  {journal} {\bibinfo  {journal} {JHEP}\
  }\textbf {\bibinfo {volume} {02}},\ \bibinfo {pages} {157} (\bibinfo {year}
  {2019})},\ \Eprint {http://arxiv.org/abs/1811.03476} {arXiv:1811.03476
  [hep-ph]} \BibitemShut {NoStop}%
%%CITATION = ARXIV:1811.03476;%%
\bibitem [{\citenamefont {Alwall}\ \emph {et~al.}(2011)\citenamefont {Alwall},
  \citenamefont {Herquet}, \citenamefont {Maltoni}, \citenamefont {Mattelaer},\
  and\ \citenamefont {Stelzer}}]{Alwall:2011uj}%
  \BibitemOpen
  \bibfield  {author} {\bibinfo {author} {\bibfnamefont {J.}~\bibnamefont
  {Alwall}}, \bibinfo {author} {\bibfnamefont {M.}~\bibnamefont {Herquet}},
  \bibinfo {author} {\bibfnamefont {F.}~\bibnamefont {Maltoni}}, \bibinfo
  {author} {\bibfnamefont {O.}~\bibnamefont {Mattelaer}}, \ and\ \bibinfo
  {author} {\bibfnamefont {T.}~\bibnamefont {Stelzer}},\ }\href {\doibase
  10.1007/JHEP06(2011)128} {\bibfield  {journal} {\bibinfo  {journal} {JHEP}\
  }\textbf {\bibinfo {volume} {06}},\ \bibinfo {pages} {128} (\bibinfo {year}
  {2011})},\ \Eprint {http://arxiv.org/abs/1106.0522} {arXiv:1106.0522
  [hep-ph]} \BibitemShut {NoStop}%
%%CITATION = ARXIV:1106.0522;%%
\bibitem [{\citenamefont {Sjostrand}\ \emph {et~al.}(2008)\citenamefont
  {Sjostrand}, \citenamefont {Mrenna},\ and\ \citenamefont
  {Skands}}]{Sjostrand:2007gs}%
  \BibitemOpen
  \bibfield  {author} {\bibinfo {author} {\bibfnamefont {T.}~\bibnamefont
  {Sjostrand}}, \bibinfo {author} {\bibfnamefont {S.}~\bibnamefont {Mrenna}}, \
  and\ \bibinfo {author} {\bibfnamefont {P.~Z.}\ \bibnamefont {Skands}},\
  }\href {\doibase 10.1016/j.cpc.2008.01.036} {\bibfield  {journal} {\bibinfo
  {journal} {Comput. Phys. Commun.}\ }\textbf {\bibinfo {volume} {178}},\
  \bibinfo {pages} {852} (\bibinfo {year} {2008})},\ \Eprint
  {http://arxiv.org/abs/0710.3820} {arXiv:0710.3820 [hep-ph]} \BibitemShut
  {NoStop}%
%%CITATION = ARXIV:0710.3820;%%
\bibitem [{\citenamefont {de~Favereau}\ \emph {et~al.}(2014)\citenamefont
  {de~Favereau}, \citenamefont {Delaere}, \citenamefont {Demin}, \citenamefont
  {Giammanco}, \citenamefont {Lemaître}, \citenamefont {Mertens},\ and\
  \citenamefont {Selvaggi}}]{deFavereau:2013fsa}%
  \BibitemOpen
  \bibfield  {author} {\bibinfo {author} {\bibfnamefont {J.}~\bibnamefont
  {de~Favereau}}, \bibinfo {author} {\bibfnamefont {C.}~\bibnamefont
  {Delaere}}, \bibinfo {author} {\bibfnamefont {P.}~\bibnamefont {Demin}},
  \bibinfo {author} {\bibfnamefont {A.}~\bibnamefont {Giammanco}}, \bibinfo
  {author} {\bibfnamefont {V.}~\bibnamefont {Lemaître}}, \bibinfo {author}
  {\bibfnamefont {A.}~\bibnamefont {Mertens}}, \ and\ \bibinfo {author}
  {\bibfnamefont {M.}~\bibnamefont {Selvaggi}} (\bibinfo {collaboration}
  {DELPHES 3}),\ }\href {\doibase 10.1007/JHEP02(2014)057} {\bibfield
  {journal} {\bibinfo  {journal} {JHEP}\ }\textbf {\bibinfo {volume} {02}},\
  \bibinfo {pages} {057} (\bibinfo {year} {2014})},\ \Eprint
  {http://arxiv.org/abs/1307.6346} {arXiv:1307.6346 [hep-ex]} \BibitemShut
  {NoStop}%
%%CITATION = ARXIV:1307.6346;%%
\bibitem [{\citenamefont {Thamm}\ \emph {et~al.}(2015)\citenamefont {Thamm},
  \citenamefont {Torre},\ and\ \citenamefont {Wulzer}}]{Thamm:2015zwa}%
  \BibitemOpen
  \bibfield  {author} {\bibinfo {author} {\bibfnamefont {A.}~\bibnamefont
  {Thamm}}, \bibinfo {author} {\bibfnamefont {R.}~\bibnamefont {Torre}}, \ and\
  \bibinfo {author} {\bibfnamefont {A.}~\bibnamefont {Wulzer}},\ }\href
  {\doibase 10.1007/JHEP07(2015)100} {\bibfield  {journal} {\bibinfo  {journal}
  {JHEP}\ }\textbf {\bibinfo {volume} {07}},\ \bibinfo {pages} {100} (\bibinfo
  {year} {2015})},\ \Eprint {http://arxiv.org/abs/1502.01701} {arXiv:1502.01701
  [hep-ph]} \BibitemShut {NoStop}%
%%CITATION = ARXIV:1502.01701;%%
\bibitem [{\citenamefont {Ball}\ \emph {et~al.}(2015)\citenamefont {Ball} \emph
  {et~al.}}]{Ball:2014uwa}%
  \BibitemOpen
  \bibfield  {author} {\bibinfo {author} {\bibfnamefont {R.~D.}\ \bibnamefont
  {Ball}} \emph {et~al.} (\bibinfo {collaboration} {NNPDF}),\ }\href {\doibase
  10.1007/JHEP04(2015)040} {\bibfield  {journal} {\bibinfo  {journal} {JHEP}\
  }\textbf {\bibinfo {volume} {04}},\ \bibinfo {pages} {040} (\bibinfo {year}
  {2015})},\ \Eprint {http://arxiv.org/abs/1410.8849} {arXiv:1410.8849
  [hep-ph]} \BibitemShut {NoStop}%
%%CITATION = ARXIV:1410.8849;%%
\bibitem [{\citenamefont {Lindner}\ \emph
  {et~al.}(2016{\natexlab{b}})\citenamefont {Lindner}, \citenamefont {Queiroz},
  \citenamefont {Rodejohann},\ and\ \citenamefont {Yaguna}}]{Lindner:2016lxq}%
  \BibitemOpen
  \bibfield  {author} {\bibinfo {author} {\bibfnamefont {M.}~\bibnamefont
  {Lindner}}, \bibinfo {author} {\bibfnamefont {F.~S.}\ \bibnamefont
  {Queiroz}}, \bibinfo {author} {\bibfnamefont {W.}~\bibnamefont {Rodejohann}},
  \ and\ \bibinfo {author} {\bibfnamefont {C.~E.}\ \bibnamefont {Yaguna}},\
  }\href {\doibase 10.1007/JHEP06(2016)140} {\bibfield  {journal} {\bibinfo
  {journal} {JHEP}\ }\textbf {\bibinfo {volume} {06}},\ \bibinfo {pages} {140}
  (\bibinfo {year} {2016}{\natexlab{b}})},\ \Eprint
  {http://arxiv.org/abs/1604.08596} {arXiv:1604.08596 [hep-ph]} \BibitemShut
  {NoStop}%
%%CITATION = ARXIV:1604.08596;%%
\bibitem [{\citenamefont {Cid~Vidal}\ \emph {et~al.}(2018)\citenamefont
  {Cid~Vidal} \emph {et~al.}}]{CidVidal:2018eel}%
  \BibitemOpen
  \bibfield  {author} {\bibinfo {author} {\bibfnamefont {X.}~\bibnamefont
  {Cid~Vidal}} \emph {et~al.} (\bibinfo {collaboration} {Working Group 3}),\
  }\href@noop {} {\  (\bibinfo {year} {2018})},\ \Eprint
  {http://arxiv.org/abs/1812.07831} {arXiv:1812.07831 [hep-ph]} \BibitemShut
  {NoStop}%
%%CITATION = ARXIV:1812.07831;%%
\bibitem [{\citenamefont {Dev}\ \emph {et~al.}(2018)\citenamefont {Dev},
  \citenamefont {Ramsey-Musolf},\ and\ \citenamefont {Zhang}}]{Dev:2018sel}%
  \BibitemOpen
  \bibfield  {author} {\bibinfo {author} {\bibfnamefont {P.~S.~B.}\
  \bibnamefont {Dev}}, \bibinfo {author} {\bibfnamefont {M.~J.}\ \bibnamefont
  {Ramsey-Musolf}}, \ and\ \bibinfo {author} {\bibfnamefont {Y.}~\bibnamefont
  {Zhang}},\ }\href {\doibase 10.1103/PhysRevD.98.055013} {\bibfield  {journal}
  {\bibinfo  {journal} {Phys. Rev.}\ }\textbf {\bibinfo {volume} {D98}},\
  \bibinfo {pages} {055013} (\bibinfo {year} {2018})},\ \Eprint
  {http://arxiv.org/abs/1806.08499} {arXiv:1806.08499 [hep-ph]} \BibitemShut
  {NoStop}%
%%CITATION = ARXIV:1806.08499;%%
\bibitem [{\citenamefont {Dev}\ \emph {et~al.}(2019)\citenamefont {Dev},
  \citenamefont {Khan}, \citenamefont {Mitra},\ and\ \citenamefont
  {Rai}}]{Dev:2019hev}%
  \BibitemOpen
  \bibfield  {author} {\bibinfo {author} {\bibfnamefont {P.~S.~B.}\
  \bibnamefont {Dev}}, \bibinfo {author} {\bibfnamefont {S.}~\bibnamefont
  {Khan}}, \bibinfo {author} {\bibfnamefont {M.}~\bibnamefont {Mitra}}, \ and\
  \bibinfo {author} {\bibfnamefont {S.~K.}\ \bibnamefont {Rai}},\ }\href
  {\doibase 10.1103/PhysRevD.99.115015} {\bibfield  {journal} {\bibinfo
  {journal} {Phys. Rev.}\ }\textbf {\bibinfo {volume} {D99}},\ \bibinfo {pages}
  {115015} (\bibinfo {year} {2019})},\ \Eprint
  {http://arxiv.org/abs/1903.01431} {arXiv:1903.01431 [hep-ph]} \BibitemShut
  {NoStop}%
%%CITATION = ARXIV:1903.01431;%%
\bibitem [{\citenamefont {Bhupal~Dev}\ \emph {et~al.}(2018)\citenamefont
  {Bhupal~Dev}, \citenamefont {Mohapatra},\ and\ \citenamefont
  {Zhang}}]{Dev:2018upe}%
  \BibitemOpen
  \bibfield  {author} {\bibinfo {author} {\bibfnamefont {P.~S.}\ \bibnamefont
  {Bhupal~Dev}}, \bibinfo {author} {\bibfnamefont {R.~N.}\ \bibnamefont
  {Mohapatra}}, \ and\ \bibinfo {author} {\bibfnamefont {Y.}~\bibnamefont
  {Zhang}},\ }\href {\doibase 10.1103/PhysRevD.98.075028} {\bibfield  {journal}
  {\bibinfo  {journal} {Phys. Rev.}\ }\textbf {\bibinfo {volume} {D98}},\
  \bibinfo {pages} {075028} (\bibinfo {year} {2018})},\ \Eprint
  {http://arxiv.org/abs/1803.11167} {arXiv:1803.11167 [hep-ph]} \BibitemShut
  {NoStop}%
%%CITATION = ARXIV:1803.11167;%%
\bibitem [{\citenamefont {Nomura}\ \emph {et~al.}(2018)\citenamefont {Nomura},
  \citenamefont {Okada},\ and\ \citenamefont {Yokoya}}]{Nomura:2017abh}%
  \BibitemOpen
  \bibfield  {author} {\bibinfo {author} {\bibfnamefont {T.}~\bibnamefont
  {Nomura}}, \bibinfo {author} {\bibfnamefont {H.}~\bibnamefont {Okada}}, \
  and\ \bibinfo {author} {\bibfnamefont {H.}~\bibnamefont {Yokoya}},\ }\href
  {\doibase 10.1016/j.nuclphysb.2018.02.011} {\bibfield  {journal} {\bibinfo
  {journal} {Nucl. Phys.}\ }\textbf {\bibinfo {volume} {B929}},\ \bibinfo
  {pages} {193} (\bibinfo {year} {2018})},\ \Eprint
  {http://arxiv.org/abs/1702.03396} {arXiv:1702.03396 [hep-ph]} \BibitemShut
  {NoStop}%
%%CITATION = ARXIV:1702.03396;%%
\bibitem [{\citenamefont {Swartz}(1989)}]{Swartz:1989qz}%
  \BibitemOpen
  \bibfield  {author} {\bibinfo {author} {\bibfnamefont {M.~L.}\ \bibnamefont
  {Swartz}},\ }\href {\doibase 10.1103/PhysRevD.40.1521} {\bibfield  {journal}
  {\bibinfo  {journal} {Phys. Rev.}\ }\textbf {\bibinfo {volume} {D40}},\
  \bibinfo {pages} {1521} (\bibinfo {year} {1989})}\BibitemShut {NoStop}%
%%CITATION = PHRVA,D40,1521;%%
\end{thebibliography}%
\end{document}